\documentclass[twocolumn,
pra
,superscriptaddress
,floatfix
,aps
,10pt
,showpacs]{revtex4-2}

\usepackage{graphicx}
\usepackage{amssymb}
\usepackage{amsmath}
\usepackage{multirow}
\usepackage{array,tabularx}
\usepackage{xcolor}
\usepackage{comment}
\usepackage[colorlinks,urlcolor=blue,citecolor=blue,linkcolor=blue]{hyperref}

\usepackage{textcomp}

\newcommand{\iac}[1]{{\color{magenta}#1}}

\usepackage{soul}
\usepackage{float}
\usepackage{upgreek}
\usepackage{makeidx}
\DeclareGraphicsExtensions{.png,.jpg,.pdf,.mps,.gif,.bmp,.eps}
\graphicspath{{figures/}}
\usepackage{physics}
\begin{document}

\title{Optical isolators based on non-reciprocal four-wave mixing}

\author{A. Mu$\tilde{\mathrm{n}}$oz de las Heras}
\email{alberto.munoz@iff.csic.es}
\affiliation{Instituto de Física Fundamental (IFF), CSIC, Calle Serrano 113b, 28006 Madrid, Spain}
\affiliation{INO-CNR BEC Center and Dipartimento di Fisica, Universit$\grave{a}$ di Trento, 38123 Trento, Italy}

\author{I. Carusotto}
\affiliation{INO-CNR BEC Center and Dipartimento di Fisica, Universit$\grave{a}$ di Trento, 38123 Trento, Italy}

\date{\today}

%
%%%%%%%%%%%%%%%%%%%%%%%%%%%%%%%%%%%%%%%%%%%%%%%%%%%%%%%%%%%%%%%%%%%%
\begin{abstract}

In this work we propose and theoretically characterize optical isolators consisting of an all-dielectric and non-magnetic resonator featuring an intensity-dependent refractive index and a strong coherent field propagating in a single direction. Such devices can be straightforwardly realized in state-of-the-art integrated photonics platforms.
The mechanism underlying optical isolation is based on the breaking of optical reciprocity induced by the asymmetric action of four-wave mixing processes coupling a strong propagating pump field with co-propagating signal/idler modes but not with reverse-propagating ones.
Taking advantage of a close analogy with fluids of light, our proposed isolation mechanism is physically understood in terms of the Bogoliubov dispersion of collective excitations on top of the strong pump beam. A few most relevant set-ups realizing our proposal are specifically investigated, such as a coherently illuminated passive ring resonator and unidirectionally lasing ring or Taiji resonators.

%Here we numerically demonstrate effective optical isolation over a broad frequency range in all-dielectric and non-magnetic devices operating at telecommunication frequencies. The proposed strategy capitalizes on the reciprocity breaking due to the asymmetric action of four-wave mixing coupling a large intensity pump with the transmitted signal but not with reverse-propagating undesired ones. This allows our devices to circumvent the dynamic reciprocity restrictions~\cite{Shi_2015}. Our theoretical formalism is based on a linearized analysis of the signal and idler fields which are treated as small perturbations in the temporal coupled-mode theory equations for the large intensity pump. Nonreciprocity is further investigated by diagonalizing the Bogoliubov matrix for fields propagating in the two directions and examining the resulting spectrum of eigenvalues and the norm of the associated eigenvectors. We propose three setups where this one scheme can be realized in different flavors: a passive ring resonator in add-drop configuration optically pump from one side, an active ring resonator in which an external bias triggers lasing in the forward direction, and an incoherently pumped active Taiji resonator.

\end{abstract}

\maketitle

%%%%%%%%%%%%%%%%%%%%%%%%%%%%%%%%%%% INTRODUCTION
%%%%%%%%%%%%%%%%%%%%%%%%%%%%%%%%%%

\section{Introduction}
\label{sec:TaijiDiode_Introduction}

Optical isolators~\cite{Jalas_2013} are devices that allow the transmission of light in a certain direction (known as \textit{forward}) while simultaneously preventing light propagation in the opposite direction (labeled \textit{reverse}). Optical isolation requires breaking Lorentz reciprocity~\cite{Potton_2004}. For this purpose many strategies have been proposed: Some authors employed magneto-optic materials that explicitly break time-reversal symmetry under a strong external magnetic field~\cite{Dotsch_2005,Bi_2011,Shoji_2012,Yan_2020}. Others relied on the application of some external drive to produce a time-dependent modulation of the material's refractive index~\cite{Bhandare05,yu2009complete,Fang_2012,Galland:13,doerr2014silicon}. An even more promising strategy to break reciprocity consists of exploiting the optical nonlinearity of the dielectric material and induce non-linear optical processes via a strong external beam. Such all-optical isolators have the important advantage of a straightforward integration in state-of-the-art silicon-based photonic networks with a small on-chip footprint.

As a first step along this line, several authors have demonstrated non-reciprocal transmission by comparing the transmitted intensity for a strong light beam injected first in the forward and then in the reverse direction~\cite{Fan_2012,Bender_2013,Peng_2014,Chang_2014,MunozDeLasHeras_2021}.
In spite of the different transmission observed in the two directions, it was pointed out in Ref.~\cite{Shi_2015} that such a process can not be considered as true optical isolation: it is in fact not generally true that the device is able to block arbitrary backwards-propagating noise when a strong signal is being transmitted in the forward direction. Such an effective blocking was highlighted only in specific cases, for instance when the frequency spectrum of the noise overlaps with that of the forward signal~\cite{Shi_2015} or when both the pump and the noise are monochromatic waves at the same frequency~\cite{DelBino_2018}.

%However, the work of Ref.~\cite{Shi_2015} showed that such systems cannot grant isolation against arbitrary backwards-propagating noise when a strong signal is being transmitted in the forward direction. This is called \textit{dynamic reciprocity}, and it constitutes an extremely important weak point as these are the usual operation conditions in which an optical isolator is expected to work.

%Nevertheless, there are some exceptions to dynamic reciprocity. Ref.~\cite{Shi_2015} demonstrated that when the frequency spectrum of the noise overlaps with that of the forward signal, four-wave mixing (FWM) coupling between the two fields preserves nonreciprocity. A specific case in this class is obtained when both signal and noise are monochromatic waves featuring the same frequency. An example is provided by Ref.~\cite{DelBino_2018}, who capitalized on the different Kerr nonlinearity shifts of the resonance frequencies of two counterpropagating whispering gallery modes in a silicon ring resonator~\cite{DelBino_2017}, namely clockwise (CW) and counterclockwise (CCW). After being transmitted through the device, light was reflected at a mirror and could not couple to the counterpropagating mode of the resonator.

In this paper we theoretically demonstrate that all-dielectric nonlinear ring resonators featuring a strong coherent field propagating in a single direction can behave as efficient optical isolators as they display a different transmittance for weak signals in the two directions: for instance, they permit an efficient transmission in the forward direction while they block transmission in the reverse one. 

This non-reciprocal behavior stems from the fact that four-wave-mixing (FWM)-induced coupling between pump and signal/idler modes is only possible when all fields propagate in the same direction and is instead strongly non-phase-matched in the reverse direction. As a result, a forward-propagating incident weak signal field automatically gives rise to a corresponding idler field at a symmetrically located frequency with respect to the pump, which allows to circumvent the restrictions imposed by dynamic reciprocity highlighted in~\cite{Shi_2015}.

To make our proposal concrete and of direct use in view of experiments, we focus our attention on three specific setups where our concept can be realized in an integrated photonics framework with state-of-the-art technology. While the three setups are based on the same microscopic mechanism, they display minor operation differences that are potentially useful for specific applications.

The first proposed setup consists of a passive ring resonator coupled to a pair of bus waveguides in add-drop configuration. The system is illuminated through one of the bus waveguides by a strong and coherent pump beam, close to resonance with a cavity mode. Generalizing the results of~\cite{DelBino_2018}, efficient isolation is demonstrated for a weak signal whose frequency spectrum is either in the vicinity of the strong pump or next to some neighboring cavity mode. In the former configuration, the coherent pump could be the output of some laser source to be injected in a photonic network, so back-propagating noise in the vicinity of the laser frequency is efficiently blocked.

The second proposed setup consists again of a ring resonator but, instead of being illuminated by an external coherent light beam, the resonator is itself endowed of optical gain and displays a single-mode laser oscillation behaviour. The coherent laser field then plays the role of the strong pump field breaking optical reciprocity and ensuring the optical isolation behavior. Since laser operation in a ring resonator would naturally occur in a randomly-chosen direction~\cite{MunozDeLasHeras_2021b}, a weak, externally injected drive through one of the bus waveguides may be needed to deterministically trigger the chirality of laser operation. 

The need for this trigger beam is removed in the third proposed setup based on a Taiji resonator (TJR) laser. In such devices, the presence of an additional S-shaped element deterministically stabilizes unidirectional laser operation via an effective nonlinear dynamical breaking of time-reversal symmetry~\cite{Hohimer_1993,Hohimer_1993b,MunozDeLasHeras_2021b} and provides a stand-alone configuration for efficient optical isolation. In order to prevent the signal beam from triggering undesired instabilities in the lasing resonator and to permit an efficient spectral rejection of the laser light from the optical network downstream, it is beneficial in this case to operate optical isolation on some neighboring Taiji resonator modes.

%nd play a central role in order to preserve the topological protection of a pair of chiral surface modes in quantum spin-Hall topological lasers~\cite{Bandres_2018,Harari_2018,Ozawa_2019,Ota_2020} due to the high robustness of their unidirectional emission against backscattering coupling light into counterpropagating modes.

Our theoretical approach is based on a time-dependent coupled mode theory. We consider the signal and idler fields as small perturbations to the strong monochromatic field in the pump mode, whose intensity is assumed to be much larger than that of any other field inside the resonator, and we linearize the motion equations with respect to signal and idler. For each configuration, we then look for the steady-state of the coupled-mode equations either under the incident coherent light or in the presence of gain, and we evaluate the frequency-dependent transmittance of a weak additional beam across the device in the two directions. To better understand the role of the asymmetry of the FWM process in the dynamics, we set up the $4\times 4$ Bogoliubov matrix connecting the signal and idler fields in the two directions, whose eigenmodes provide direct information on the position of the transmittance peaks and on the dynamical stability of the system. This Bogoliubov picture reveals a direct connection between our optical isolator device and the collective dynamics of flowing polariton condensates under either a coherent or an incoherent pumping~\cite{Carusotto_2013}.

The article is organized as follows: Sec.~\ref{sec:TaijiDiode_TheoreticalModel} introduces the coupled-mode theory and the linearized approximation employed in our calculation. Our results for the case of a passive nonlinear ring resonator coherently pumped from one side are shown in Sec.~\ref{sec:TaijiDiode_PassiveRing}. Sec.~\ref{sec:TaijiDiode_ActiveRing} is then devoted to the case of active ring resonators in a lasing regime. Our results for the TJR laser isolator are reported in Sec.~\ref{sec:TaijiDiode_TJRlaser}. Conclusions are finally drawn in Sec.~\ref{sec:Conclusions}.

\begin{figure*}[t]
    \centering
    \includegraphics[width=\textwidth]{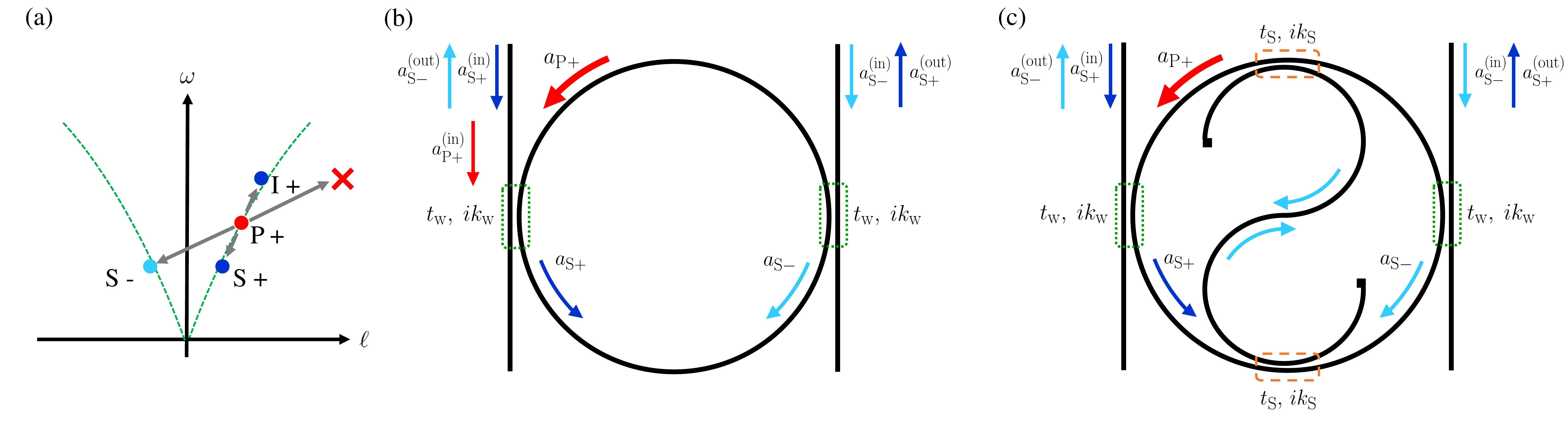}
    \caption{\textbf{(a)} Dispersion relation of the resonant frequencies $\omega_\ell^{(0)}$ of a ring or Taiji (TJR) resonator as a function of the angular momentum $\ell$ of the mode. The green dashed lines are the dispersion relations for CCW ($\ell>0$) and CW ($\ell<0$) modes including the curvature given by Eq.~\eqref{eq:DispersionRelationLambda}. While FWM (symbolized by the gray arrows) can effectively couple CCW-propagating pump, signal and idler modes, it cannot couple the pump P$+$ with the signal S$-$ because the resulting idler would completely fall out of resonance. \textbf{(b)} Scheme of a ring resonator pumped in the CCW direction by an external field $a^{\rm (in)}_{\rm P+}$. The dashed green rectangles indicate directional couplers coupling the resonator to the bus waveguides in the add-drop configuration, with transmission and coupling amplitudes $t_{\rm w}$ and $ik_{\rm w}$, respectively. The field amplitude $a_{\rm P+}$ corresponds to the large intensity pump mode inside the resonator. The system is then probed in the two opposite directions by the external fields $a^{\rm (in)}_{\rm S\pm}$ (blue and cyan arrows). The field amplitudes of these modes inside the resonator are given by $a_{\rm S\pm}$. \textbf{(c)} Similar scheme for a TJR in which light is unidirectionally coupled from the CW into the CCW direction by the S-shaped waveguide. The dashed red rectangles indicate directional couplers coupling the ring resonator to the S-shaped element, with transmission and coupling amplitudes $t_{\rm S}$ and $ik_{\rm S}$, respectively.
    % The transmission and coupling parameters for the directional couplers between the ring and the S waveguides are given by $t_{\rm S}$ and $ik_{\rm S}$, respectively.
    }
    \label{fig:DiodeDiagram}
\end{figure*}
%

%%%%%%%%%%%%%%%%%%%%%%%%%%%%%%%%%%% THEORETICAL MODEL
%%%%%%%%%%%%%%%%%%%%%%%%%%%%%%%%%%

\section{The physical system and the theoretical model}
\label{sec:TaijiDiode_TheoreticalModel}

In this first Section we introduce the device and the optical configuration under investigation and we present the theoretical framework based on temporal coupled-mode equations that we are going to use in the following Sections to describe its optical response in the different cases.

%\subsection{Resonant modes of a ring resonator}

\subsection{The physical system}

The specific setup we are going to consider in this work consists of a standard ring resonator (see Fig.~\ref{fig:DiodeDiagram}b) or a Taiji resonator (TJR, see Fig.~\ref{fig:DiodeDiagram}c) coupled to a pair of bus waveguides located at opposite sides in the so-called \textit{add-drop} configuration. Light is injected ({\rm in}) into the resonator through one of the bus waveguides and transmittance ({\rm out}) is then measured at the output of the other bus waveguide.

Light propagating inside a ring resonator of radius $R$ and linear refractive index $n_{\rm L}$ is characterized by a series of discrete modes labelled by an integer number $\ell$ that determines the angular momentum of the mode. The sign of $\ell$ describes its propagation direction: $\ell>0$ indicates a counter-clockwise (CCW)-propagating mode, while $\ell<0$ corresponds to a clockwise (CW)-propagating one. 
In the linear optical regime, the resonance frequencies are given by %respectively given by}
%
% \begin{align}
%     \lambda^{(0)}_{\ell}=\frac{2\pi R n_{\rm L}(\ell)}{|\ell|},\;\;\;
\begin{equation}    \omega^{(0)}_{\ell}=\frac{c}{n_{\rm L}(\ell)}\frac{|\ell|}{R}\, ,
\label{eq:lambda_l} \end{equation}
%\end{align}
%
where $c$ is the vacuum speed of light.
Since resonators are typically built of a dispersive dielectric medium, the linear refractive index $n_{\rm L}$ depends on the angular momentum $\ell$ of the mode via the mode frequency. We can consider the Taylor expansion of $n_{\rm L}(\ell)$ around a certain angular momentum mode, which we label $\ell_{\rm P}$:
\begin{align}
    n_{\rm L}(\ell)=n_{\rm L}(|\ell_{\rm P}|)+\left(\frac{dn_{\rm L}}{d|\ell|}\right)_{|\ell_{\rm P}|}(|\ell|-|\ell_{\rm P}|)+\mathcal{O}(\ell^2).
\label{eq:Taylor_n_l}
\end{align}
By combining Eqs.~(\ref{eq:lambda_l}) and (\ref{eq:Taylor_n_l}), we arrive at the expression
%
%\begin{align}
%    \omega^{(0)}_{\ell} &=\frac{c}{n_{\rm L}(|\ell_{\rm P}|)}
%    \left[1+\frac{1}{n_{\rm L}(|\ell_{\rm P}|)}\left(\frac{dn_{\rm L}}{d|\ell|}\right)_{|\ell_{\rm P}|}|\ell_{\rm P}|\right]\frac{|\ell|}{R}
%    \nonumber\\
%    & -\frac{c}{n^2_{\rm L}(|\ell_{\rm P}|)}\left(\frac{dn_{\rm L}}{d|\ell|}\right)_{|\ell_{\rm P}|}\frac{\ell^2}{R},
%\label{eq:DispersionRelation1}
%\end{align}
%
%We now capitalize on Eq.~\eqref{eq:lambda_l}
%in order to rewrite Eq.~\eqref{eq:DispersionRelation1} in terms of the derivative $dn_{\rm L}/d\lambda$:
%
%\begin{align}
%    \omega^{(0)}_{\ell} &=\frac{c}{|\ell_{\rm P}|}\left[1-\frac{2\pi R}{|\ell_{\rm P}|}\left(\frac{dn_{\rm L}}{d\lambda}\right)_{\lambda_{\rm P}}\right]\frac{|\ell|}{R}
%    \nonumber\\
%    &+\frac{c}{n_{\rm L}(|\ell_{\rm P}|)}\frac{2\pi}{\ell^2_{\rm P}}\left(\frac{dn_{\rm L}}{d\lambda}\right)_{\lambda_{\rm P}}\ell^2,
%\label{eq:DispersionRelationLambda_prev}
%\end{align}
%
%
\begin{align}
    \omega^{(0)}_{\ell} &=\omega^{(0)}_{\rm P}+v\left( |\ell|-|\ell_{\rm P}| \right)+\frac{\alpha}{2}\left( |\ell|-|\ell_{\rm P}| \right)^2,
\label{eq:DispersionRelationLambda}
\end{align}
for the $\ell$-dependent dispersion relation of light in the ring resonator, where we have defined
\begin{align}
    \omega^{(0)}_{\rm P}&\equiv\omega^{(0)}_{\ell_{\rm P}} =\frac{c|\ell_{\rm P}|}{n_{\rm L}(|\ell_{\rm P}|)R}
    ,\label{eq:omega0} \\
    v & = \frac{c}{n_{\rm L}(|\ell_{\rm P}|)R}\left[1+\frac{2\pi R}{|\ell_{\rm P}|}\left(\frac{dn_{\rm L}}{d\lambda}\right)_{\lambda^{(0)}_{\rm P}}\right]
    ,\\
    \alpha & = 2 \frac{c}{n_{\rm L}(|\ell_{\rm P})|}\frac{2\pi}{\ell^2_{\rm P}} \left(\frac{dn_{\rm L}}{d\lambda}\right)_{\lambda^{(0)}_{\rm P}}, \label{eq:alpha}
\end{align}
%
%Above we use the simplified notation $\omega^{(0)}_{\rm P}\equiv\omega^{(0)}_{\ell_{\rm P}}$.
and the derivatives are evaluated at $\lambda^{(0)}_{\rm P}=2\pi R n_{\rm L}/|\ell_{\rm P}|$. The second-order expansion in Eq.~\eqref{eq:DispersionRelationLambda} with the coefficients in Eqs.(\ref{eq:omega0}-\ref{eq:alpha}) is valid under the quite weak condition
\begin{align}
    \left(\frac{dn_{\rm L}}{d\lambda}\right)_{\lambda^{(0)}_{\rm P}}
    \gg
    \left(\frac{d^2n_{\rm L}}{d\lambda^2}\right)_{\lambda^{(0)}_{\rm P}}\lambda^{(0)}_{\rm P}\,:
\label{eq:ConditionQuadraticDispersion}
\end{align}
which is well satisfied by typical integrated photonics devices. For instance, SiON waveguides of typical transverse area $1200$ nm $\times$ $570$ nm  operating around $\lambda^{(0)}_{\rm P}=1500$ nm display an effective linear refractive index (including the effect of confinement) $n_{\rm L}\simeq 1.59$, while the first derivative is $(dn_{\rm L}/d\lambda)_{\lambda^{(0)}_{\rm P}}=-1.42\times 10^{-4}$ nm$^{-1}$ and the second derivative is zero with a large precision~\cite{MunozDeLasHeras_2021}.
%: as a result, the condition~\eqref{eq:ConditionQuadraticDispersion} is well fulfilled and the quadratic approximation for the dispersion relation is valid.
A sketch of the dispersion relation given by Eq.~\eqref{eq:DispersionRelationLambda} can be found in Fig.~\ref{fig:DiodeDiagram}a.

%\subsection{The theoretical model}

The key idea behind our proposal for optical isolators is to have the resonator filled with a strong intensity, propagating field at a frequency $\omega_{\rm P}$ in the vicinity of the linear resonance frequency $\omega^{(0)}_{\rm P}$ of a mode of angular momentum $\ell_{\rm P}$, indicated in the following as P+ mode. Without loss of generality we take the propagation to be in the CCW direction with a positive angular momentum $\ell_{\rm P}>0$. 
Thanks to the intrinsic optical nonlinearity of the material, a pair of photons from the P$+$ mode can scatter into the S$+$ (signal) and I$+$ (idler) modes of angular momenta $\ell_{\rm S}$ and $\ell_{\rm I} = 2\ell_{\rm P}-\ell_{\rm S}$ and resonance frequencies  $\omega^{(0)}_{\rm S,I}\equiv\omega^{(0)}_{\ell_{\rm S,I}}$. % and \iac{$\omega^{(0)}_{\rm I}\equiv\omega^{(0)}_{\ell_{\rm I}}$}.
%($\omega_{\rm I}=2\omega_{\rm P}-\omega_{\rm S}$), 
Such a process is known as \textit{four-wave mixing} (FWM).
%\iac{In view of the following developments, it is important to note that the wavelength-dependence of the refractive index $n_{\rm L}(\lambda)$ and the frequency detuning of the pump $\omega_P-\omega_P^{(0)}$ make the S$+$ and I$+$ modes to be detuned with respect to the corresponding resonance frequencies $\omega^{(0)}_{\rm S}\equiv\omega^{(0)}_{\ell_{\rm S}}$ and $\omega^{(0)}_{\rm I}\equiv\omega^{(0)}_{\ell_{\rm I}}$ given by the dispersion relation~\eqref{eq:DispersionRelationLambda}. 
%Furthermore, as we will see, when the optical nonlinearity of the material is considered the resonance frequencies $\omega^{(0)}_{\rm P,S,I}$ of pump, signal, and idler will also be shifted with respect to the linear value appearing in Eq.~\eqref{eq:DispersionRelationLambda}. Such a deviation is proportional to the pump intensity.
The non-reciprocal behavior is then probed by comparing the transmittance of weak signal beams incident from opposite directions on the resonator around the $\omega^{(0)}_{\rm S}$ resonance, as indicated by the blue and cyan arrows in Fig.~\ref{fig:DiodeDiagram}b. When the signal is coupled into the S$+$ mode co-propagating with the pump in the CCW direction, FWM processes are able to significantly alter its transmittance spectrum. On the other hand, a signal propagating in the CW direction in the S$-$ mode is not coupled to the pump P$+$ through by FWM processes since no available resonance exists in the vicinity of the resulting I$-$ mode determined by conservation of energy and angular momentum (see the sketch in Fig.~\ref{fig:DiodeDiagram}a). Such an asymmetrical behaviour of the FWM processes for signal beams propagating in opposite directions is responsible for the breaking of reciprocity and leads to the optical isolator behavior: in what follows, we set the pump direction as the \textit{forward} operation direction (in our case we chose the CCW) for the optical isolator, and its opposite direction as the \textit{reverse} one (in our case the CW).

\subsection{The theoretical model}

Within a standard coupled-mode theory approach~\cite{butcher_cotter_1990}, the partial differential equation describing the temporal evolution of the complete field amplitude $a(\varphi,t)$ at the angular position $\varphi$ along the ring (see Appendix~\ref{sec:Appendix}) can be projected into a set of ordinary differential equations for the different modes. These constitute the central tool in our theoretical description.

In the simplest pump-only configuration $a(\varphi,t)=\tilde{a}_{\rm P+}(t)\,e^{i\ell_{\rm P} \varphi}\,e^{-i\omega_{\rm P} t}$ and the evolution equation for the amplitude in the P$+$ mode, seen from the rotating frame at $\omega_{\rm P}$, has the form~\cite{Walls1994_InputOutput}
\begin{align}
    &i\frac{\partial \tilde{a}_{\rm P +}}{\partial t} =
    (\omega^{(0)}_{\rm P}-\omega_{\rm P}) \tilde{a}_{\rm P +}
    -g_{\rm NL}
    |\tilde{a}_{\rm P +}|^2
    \tilde{a}_{\rm P +}
    \nonumber\\
    &-i\gamma_{\rm T}\tilde{a}_{\rm P +}+i\frac{P_{0}}{1+\frac{1}{n_{\rm S}}|\tilde{a}_{\rm P +}|^2}\tilde{a}_{\rm P +}
    -\frac{c}{Ln_{\rm L}}\kappa_{\rm w}\tilde{a}^{\rm (in)}_{\rm P +},
\label{eq:aP}
\end{align}
for both passive and active ring resonators and TJRs. Here, $g_{\rm NL}\simeq n_{\rm NL} \omega^{(0)}_{\rm P}/n_{\rm L}$ %$g_{\rm NL}$ % = n_{\rm NL}\omega^{(0)}_{\rm P}/n_{\rm L}$ 
is the effective nonlinear coefficient and $|a_{\rm P +}|^2$ quantifies the light intensity in the P$+$ mode. For simplicity, we have assumed a local and instantaneous Kerr nonlinearity corresponding to a nonlinear refractive index $n_{\rm NL}$, which is an experimentally accessible quantity with units of inverse intensity~\cite{Trenti_2018}.

The total loss rate $\gamma_{\rm T}=\gamma_{\rm A}+\gamma_{\rm w}+\gamma_{\rm S}$ includes radiative decay into the bus waveguides $\gamma_{\rm w}$ and into the directional S-shaped coupler $\gamma_{\rm S}$ in the TJR case, as well as absorption losses $\gamma_{\rm A}$. As usual, the radiative decay rates can be related to the corresponding coupling amplitudes at the directional coupler elements by $\gamma_{\rm w,S}=ck_{\rm w,S}^2/(Ln_{\rm L})$. For convenience, an analogous amplitude $k_{\rm A}$ is introduced for the non-radiative decay $\gamma_{\rm A}$.
Gain is modelled as a local and temporally instantaneous saturable gain term~\cite{Hambenne_1975} of bare rate $P_{0}$ and saturation coefficient $n_{\rm S}$, which physically corresponds a class-A laser medium whose dynamics is fast enough to be integrated out.
The last term of Eq.~\eqref{eq:aP} represents driving of the resonator via an external driving field %$a^{\rm (in)}_{\rm P+}=\tilde{a}^{\rm (in)}_{\rm P\pm}e^{i(\ell_{\rm P}z/R-\omega_{\rm P}t)}$ 
$a^{\rm (in)}_{\rm P+}(t)=\tilde{a}^{\rm (in)}_{\rm P\pm}e^{-i\omega_{\rm P} t}$ at a frequency $\omega_{\rm P}$.  %In the following, this term will account for either the coherent pump of a passive ring resonator or for the small driving setting the lasing direction of the active ring resonator. 

The steady-state of Eq.~\ref{eq:aP} will determine the amplitude of the continuous-wave, strong field in the pump mode: under a coherent pump, this will be set by the interplay of coherent drive, nonlinearity and losses and will display various optical bistability and limiting behaviours depending on the detuning of the pump frequency $\omega_{\rm P}$ from the resonant mode frequency $\omega_{\rm P}^{(0)}$. In the case of lasing under an incoherent pump, the frequency $\omega_{\rm P}$ will be instead determined by the nonlinear frequency shift of the pump mode. These different cases will be the subject of the next Sections.

%\iac{In order to Let us express the spatio-temporal dependence of the pump field amplitude in the ring resonator in terms of a slowly varying amplitude with respect to the angular momentum $\ell_{\rm P}>0$ and the frequency $\omega_{\rm P}$, $a_{\rm P +}(z,t)=\tilde{a}_{\rm P +}(t)e^{i(\ell_{\rm P}z/R-\omega_{\rm P}t)}$, where $z$ is the spatial coordinate around the ring resonator perimeter.} 

Under the assumption of a weak incident signal beam, the signal and idler  fields can be introduced as small perturbations on top of the pump-only field in the Ansatz
\begin{multline}
    a(\varphi,t)=\tilde{a}_{\rm P+}\,e^{i(\ell_{\rm P} \varphi-i\omega_{\rm P} t)}+\\ +\tilde{a}_{\rm S+}(t)\,e^{i(\ell_{\rm S} \varphi-\omega_{\rm S} t)}+\tilde{a}_{\rm I+}\,e^{i(\ell_{\rm I} \varphi-\omega_{\rm I} t)}+ \\ +\tilde{a}_{\rm S-}\,e^{-i(\ell_{\rm S} \varphi+\omega_{\rm S} t)}+\tilde{a}_{\rm I-}(t)\,e^{-i(\ell_{\rm I} \varphi+\omega_{\rm I} t)}\,, \label{eq:ansatz}
\end{multline}
where $\tilde{a}_{\rm S\pm}$ and $\tilde{a}_{\rm I\pm}$ are the signal and idler fields in the CCW and CW directions, as seen from the rotating frame at $\omega_{\rm S,I}$, and their dynamics can be described within a linearized theory.  
% to the pump with frequency (angular momentum) $\omega_{\rm S,I}$ ($\pm|\ell_{\rm S,I}|$), respectively, i.e.
% %
% \begin{align}
%     &a_{\rm P +} \rightarrow \tilde{a}_{\rm P +}e^{i\ell_{\rm P}z/R-i\omega_{\rm P}t} 
%     \nonumber\\
%     &+
%     \tilde{a}_{\rm S +}e^{i \ell_{\rm S} z/R-i\omega_{\rm S}t)} + \tilde{a}_{\rm I +}e^{i\ell_{\rm I}z/R-i\omega_{\rm I}t}
%     \nonumber\\
%     &+
%     \tilde{a}_{\rm S -}e^{-i\ell_{\rm S} z/R-i\omega_{\rm S}t} + \tilde{a}_{\rm I -}e^{-i \ell_{\rm I}z/R - \omega_{\rm I}t}
%     .
% \label{eq:Substitution_aP}
% \end{align}
% %
% Making this replacement into Eq.~\eqref{eq:aP}, staying at linear order $\mathcal{O}(a_{\rm S \pm}, a_{\rm I \pm})$ in the perturbations,
Assuming from the outset that no pump is present in the counterpropagating P- mode (i.e. $\tilde{a}_{\rm P -}=0$) and neglecting terms representing nonlinear processes that do not conserve energy or angular momentum, the temporal coupled-mode equations for the amplitudes of pump, signal, and idler inside the resonator have the form:
\begin{widetext}
\begin{align}
    \label{eq:GeneralP}
    i\dot{\tilde{a}}_{\rm P +} &=
    (\omega^{(0)}_{\rm P}-\omega_{\rm P})\tilde{a}_{\rm P +}
    -i\gamma_{\rm T}\tilde{a}_{\rm P +}
    -\frac{c}{Ln_{\rm L}}\kappa_{\rm w}\tilde{a}^{\rm (in)}_{\rm P +}-g_{\rm NL}
    |\tilde{a}_{\rm P +}|^2
    \tilde{a}_{\rm P +}
    +i\frac{P_{0}}{1+\frac{1}{n_{\rm S}}|\tilde{a}_{\rm P +}|^2}\tilde{a}_{\rm P +},
    \\
    \label{eq:GeneralS}
    i\dot{\tilde{a}}_{\rm S \pm} &=
    (\omega^{(0)}_{\rm S}-\omega_{\rm S})\tilde{a}_{\rm S \pm}-i\gamma_{\rm T}\tilde{a}_{\rm S \pm}+\beta^{\rm (S)}_{\pm,\mp}\tilde{a}_{\rm S \mp}-\frac{c}{Ln_{\rm L}}\kappa_{\rm w}\tilde{a}^{\rm (in)}_{\rm S \pm}-2g_{\rm NL}|\tilde{a}_{\rm P +}|^2\tilde{a}_{\rm S \pm}
    -g_{\rm NL}\delta_{\pm,+}\,\tilde{a}^2_{\rm P +}\tilde{a}^{*}_{\rm I \pm}
    \nonumber\\
    &
    +i\frac{P_{0}}{1+\frac{1}{n_{\rm S}}|\tilde{a}_{\rm P +}|^2}\tilde{a}_{\rm S \pm}
    -i\frac{P_{0}/n_{\rm S}}{\left[1+\frac{1}{n_{\rm S}}|\tilde{a}_{\rm P +}|^2\right]^2}\left(|\tilde{a}_{\rm P +}|^2\tilde{a}_{\rm S \pm}+\delta_{\pm,+}\,\tilde{a}^2_{\rm P +}\tilde{a}^{*}_{\rm I \pm}\right)
    \\
    i\dot{\tilde{a}}_{\rm I \pm} &=
    (\omega^{(0)}_{\rm I}-\omega_{\rm I})\tilde{a}_{\rm I \pm}-i\gamma_{\rm T}\tilde{a}_{\rm I \pm}+\beta^{\rm (I)}_{\pm,\mp}\tilde{a}_{\rm I \mp}
    -2g_{\rm NL}|\tilde{a}_{\rm P +}|^2\tilde{a}_{\rm I \pm}
    -g_{\rm NL}\delta_{\pm,+}\tilde{a}^2_{\rm P +}\tilde{a}^{*}_{\rm S \pm}
    \nonumber\\
    &+i\frac{P_{0}}{1+\frac{1}{n_{\rm S}}|\tilde{a}_{\rm P +}|^2}\tilde{a}_{\rm I \pm}
    -i\frac{P_{0}/n_{\rm S}}{\left[1+\frac{1}{n_{\rm S}}|\tilde{a}_{\rm P +}|^2\right]^2}\left(|\tilde{a}_{\rm P +}|^2\tilde{a}_{\rm I \pm}+\delta_{\pm,+}\tilde{a}^2_{\rm P \iac{+}}\tilde{a}^{*}_{\rm S \pm}\right)
    ,
\label{eq:GeneralI}
\end{align}
\end{widetext}
where we have defined $\delta_{\pm,+}$ as a shorthand giving 1 on the $+$ mode and 0 on the $-$ one. As mentioned above, the frequency $\omega_{\rm P}$ of the pump field is set here either by the coherent pump field or by the lasing frequency. On the other hand, the frequencies of the signal and idler fields are directly set by the coherent field of amplitude $\tilde{a}^{\rm (in)}_{\rm S\pm}$ driving the signal mode at frequency $\omega_{\rm S}$ and by the frequency $\omega_{\rm I}=2\omega_{\rm P}-\omega_{\rm S}$ of the idler beam generated by FWM processes.

Two main effects of the Kerr nonlinearity proportional to $g_{\rm NL}$ can be recognized in these equations of motion~\footnote{In these equations, we have neglected for simplicity the (typically small) frequency dependence of the nonlinearity and of the radiative couplings. Under this approximation, we have taken constant values for $g_{\rm NL}=n_{\rm NL}\omega^{(0)}_{\rm P}/n_{\rm L}$ and for all $\gamma$'s regardless of the considered mode.}. The first nonlinear term is the usual frequency shift of the signal/idler mode frequencies under the effect of the Kerr optical nonlinearity, proportional to the light intensity in the strong pump mode. Compared to the analogous term in Eq.~\ref{eq:GeneralP}, the factor $2$ appearing in Eqs.~(\ref{eq:GeneralS}-\ref{eq:GeneralI}) accounts for the bosonic exchange factor in the interaction between different modes.  This nonlinear frequency shift occurs identically for both CCW and CW signal and idler modes.
The following nonlinear terms in Eqs.~(\ref{eq:GeneralS}-\ref{eq:GeneralI}) describe instead the FWM processes induced by parametric scattering processes mediated by the Kerr nonlinearity, with a strength proportional to the squared pump amplitude. Because of the angular momentum and energy conservation arguments mentioned above and sketched in Fig.~\ref{fig:DiodeDiagram}a, these FWM process are only active in the CCW signal/idler modes co-propagating with the pump field: as already mentioned, this is the key feature underlying the nonlinear breaking of reciprocity and, then, the optical isolation behavior.
%and to the complex conjugate of the amplitude $\tilde{a}^{2}_{\rm P\pm}\tilde{a}^{*}_{\rm I\pm}$ and $\tilde{a}^{2}_{\rm P\pm}\tilde{a}^{*}_{\rm S\pm}$ in each case. 
Quite interestingly, Eqs.~(\ref{eq:GeneralS}-\ref{eq:GeneralI}) also display FWM processes induced by the intrinsic nonlinearity of the saturable gain. Their magnitude, as we are going to see in what follows, is however typically much smaller than the one of the FWM processes induced by the Kerr nonlinearity and they give a negligible contribution to the non-reciprocity.

Finally, the terms proportional to $\beta^{\rm (S,I)}_{\pm,\mp}$ describe the coupling between counterpropagating modes. In the case of a simple ring resonator of Fig.\ref{fig:DiodeDiagram}b, these terms obviously vanish identically, $\beta^{\rm (S,I)}_{\pm,\mp}=0$. 
On the other hand, significant such couplings are induced by the S-shaped element of the TJR resonator sketched in Fig.\ref{fig:DiodeDiagram}c: for the chosen spatial orientation of the S-shaped element, the coupling occurs unidirectionally from the CW into the CCW direction and can be described by $\beta^{\rm (S,I)}_{\mp}=0$ and $\beta^{\rm (S,I)}_{\pm}=-i2\gamma_{\rm S}e^{i\omega_{\rm S,I} n_{\rm L}L_{\rm S}/c}$~\cite{MunozDeLasHeras_2021b}. The fact that no such term appears in the equation of motion for $\tilde{a}_{\rm P+}$ stems from the fact that lasing in a TJR always occurs in the CCW pump mode P+ which is immune to the coupling induced by the S-shaped element~\cite{MunozDeLasHeras_2021b}. The situation would be of course different in the case of a coherently pumped TJR considered in~\cite{MunozDeLasHeras_2020}, but we are not considering this case in this work.

% for signal (S) and idler (I). For a ring resonator these are $\beta^{\rm (S,I)}_{\pm,\mp}=0$, while in the case of a TJR featuring an S element of length $L_{\rm S}$ that couples light from the CW into the CCW direction one has $\beta^{\rm (S,I)}_{\mp}=0$ and $\beta^{\rm (S,I)}_{\pm}=-i2ck^{2}_{\rm S}e^{i\omega^{(0)}_{\rm S,I} n_{\rm L}L_{\rm S}/c}/L n_{\rm L}$~\cite{MunozDeLasHeras_2021b} (nonlinear effects can be safely neglected in the exponential factor).

% Overall, the linearity of the coupled-mode equations for signal~\eqref{eq:GeneralS} and idler~\eqref{eq:GeneralI} in $\tilde{a}_{\rm S\pm}$ and $\tilde{a}_{\rm I\pm}$ will be of crucial importance for demonstrating the isolator character of our devices. 
% %The robustness of the solutions under such a linearized approximation with respect to a growing signal intensity is the subject of Appendix~\ref{sec:AppStability}.

In the following Sections we solve Eqs.~(\ref{eq:GeneralP}-\ref{eq:GeneralI}) for the steady state of the field amplitude $\tilde{a}^{(0)}_{\rm P,S,I\pm}=\tilde{a}_{\rm P,S,I\pm}(t\rightarrow\infty)$ in the pump, signal, and idler modes using a 4th order Runge-Kutta algorithm. As typical parameters for a realistic silicon photonics implementation we choose $R=20$ $\mu$m, $n_{\rm L}=1.59$, $\gamma_{\rm A}n_{\rm L}/c=8\times 10^{-6}$ $\mu$m$^{-1}$ (which corresponds to $\gamma_{\rm A}=1.5$ GHz), %implies $k_{\rm A}=0.032$), 
and $\omega^{(0)}_{\rm P}=2\pi\times 200$ THz, corresponding to the mode $\ell_{\rm P}=133$, for which we can employ the value of the derivative $(dn_{\rm L}/d\lambda)_{\lambda_{\rm P}}=-1.42\times 10^{-4}$ nm$^{-1}$ presented at the beginning of this Section. These values yield $v=8.20$ THz and $\alpha=-18.8$ GHz for the dispersion relation~\eqref{eq:DispersionRelationLambda}.

Our main observable quantities will consist of the forward and reverse normalized transmittances $T_{\rm for}$ and $T_{\rm rev}$ for the signal exiting the system from the bus-waveguide opposite to that in which light is injected. In a concrete experiment, the fact that pump, signal and idler feature different frequencies is of crucial importance, as it allows to use a frequency filter at the output of the bus waveguide to isolate the signal transmittance of interest for optical isolation. Within the coupled mode theory, the transmittances are defined as usual as
\begin{align}
\label{eq:T_for}
    & T_{\rm for}=\frac{|\tilde{a}^{\rm (out)}_{\rm S+}|^2}{|\tilde{a}^{\rm (in)}_{\rm S+}|^2}=\frac{|ik_{\rm w}\tilde{a}^{(0)}_{+}|^2}{|\tilde{a}^{\rm (in)}_{\rm S+}|^2},
    \\
    & T_{\rm rev}=\frac{|\tilde{a}^{\rm (out)}_{\rm S-}|^2}{|\tilde{a}^{\rm (in)}_{\rm S-}|^2}=\frac{|ik_{\rm w}\tilde{a}^{(0)}_{-}|^2}{|\tilde{a}^{\rm (in)}_{\rm S-}|^2}\,:
\label{eq:T_rev}
\end{align}
at the level of our linearized theory, these quantities are independent of the input signal intensity. Of course, this result only holds for sufficiently weak signal intensities, which is a key assumption underlying our proposal.

In the following of this work, we are going to present our results for the transmittance spectrum of forward and reverse-propagating signals as a function of the signal frequency $\omega_{\rm S}$. The different Sections will be devoted to the different cases of coherent vs. incoherent pump and of ring vs. TJR geometries: for each of them, the advantages and disadvantages in view of optical isolation will be highlighted.

%%%%%%%%%%%%%%%%%%%%%%%%%%%%%%%%%%% PUMPED PASSIVE RING RESONATOR
%%%%%%%%%%%%%%%%%%%%%%%%%%%%%%%%%%

\section{Passive ring resonator}
\label{sec:TaijiDiode_PassiveRing}

\begin{figure}
    \centering
    \includegraphics[width=0.5\textwidth]{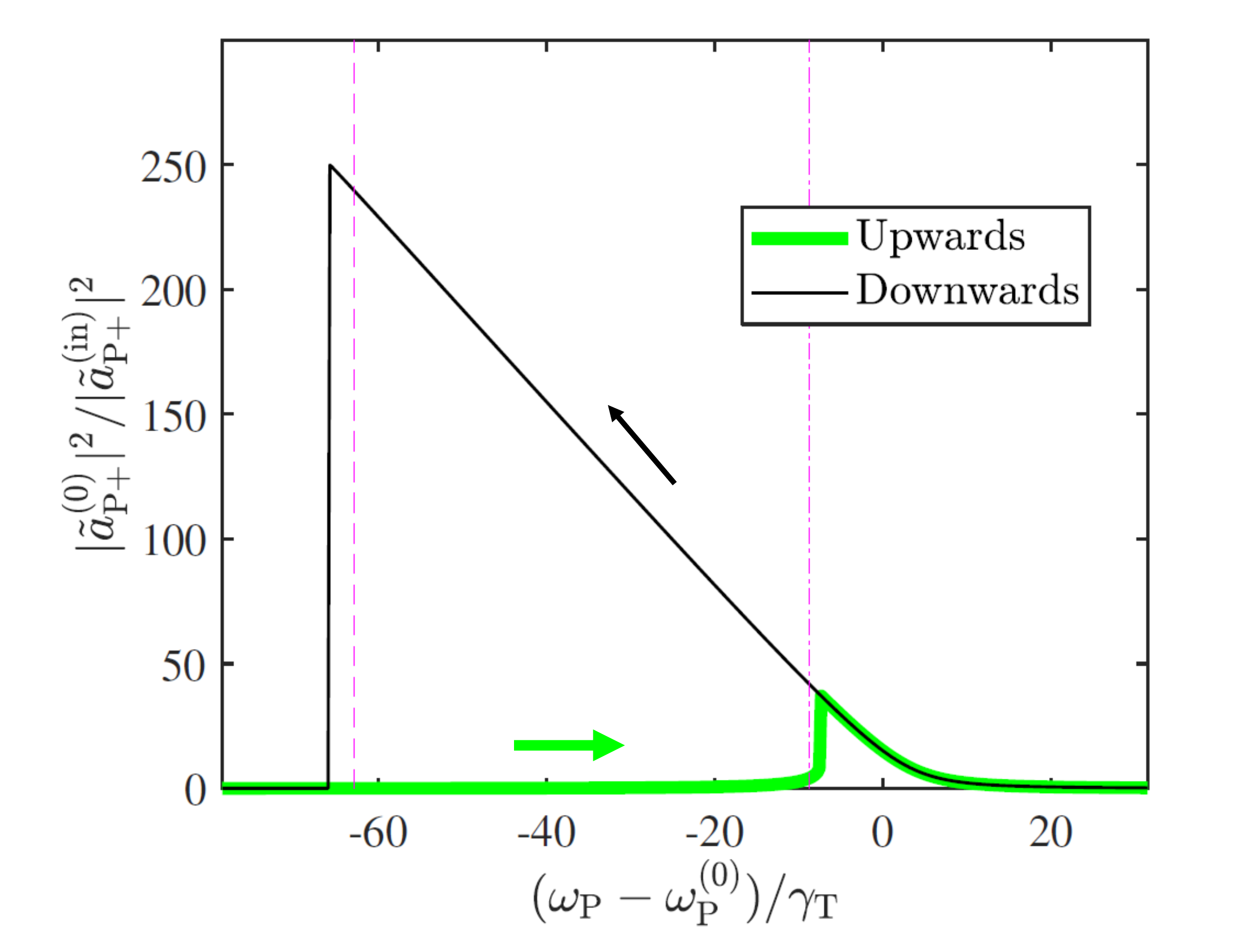}
    \caption{Optical bistability in the coherently illuminated passive ring resonator considered in Sec.~\ref{sec:TaijiDiode_PassiveRing}. The figure shows the pump intensity  $|\tilde{a}^{(0)}_{\rm P+}|^2$ circulating in the $\ell_{\rm P}=133$ mode of the resonator in units of the incident pump intensity $|\tilde{a}^{\rm (in)}_{\rm P+}|^2$, as a function of the incident frequency $\omega_{\rm P}$. When $\omega_{\rm P}$ is scanned upwards, the intensity follows the thick green curve (lower branch). In a downwards frequency ramp, the intensity follows instead the thin black line (upper branch). The vertical dashed and dash-dotted linen indicate the frequencies $\omega_{\rm P}-\omega^{(0)}_{\rm P}\simeq -63\gamma_{\rm T}$ and $\omega_{\rm P}-\omega^{(0)}_{\rm P}\simeq -8.7\gamma_{\rm T}$ used for the upper and lower branch configurations in the following Figures.}
    \label{fig:DiodeBistability}
\end{figure}

As a first example, in this Section we theoretically demonstrate non-reciprocal behavior for signal light in a coherently pumped passive ring resonator. No saturable gain is present, but the resonator is endowed of a sizable Kerr nonlinearity leading to an intensity-dependent refractive index as well as direction-dependent FWM processes.

The coherent pump of frequency $\omega_{\rm P}$ and incident intensity $|\tilde{a}^{\rm (in)}_{\rm P+}|^2$ is injected through the left-hand waveguide and sets the CCW mode as the forward operation direction. Due to the optical nonlinearity, the effective resonance frequency of the mode (given in the linear regime by Eq.~\eqref{eq:DispersionRelationLambda}) gets shifted. For the chosen value $g_{\rm NL}>0$, this shift is in the red-wards direction. As it is shown in Fig.~\ref{fig:DiodeBistability}, the pump intensity inside the resonator $|\tilde{a}^{(0)}_{\rm P+}|^2$ describes a hysteresis cycle as a function of $\omega_{\rm P}$ known as optical bistability~\cite{Gibbs_1976}, displaying the two intensity branches: for a given incident intensity $|\tilde{a}^{\rm (in)}_{\rm P+}|^2$, the internal intensity $|\tilde{a}^{(0)}_{\rm P+}|^2$ will depend on whether we are sitting on the upper or lower branch of this hysteresis cycle. An appropriate choice of initial conditions allows us to pump in the branch of interest: Sec.~\ref{sec:Passive_Upper} explores the situation in which the pump is located in the upper bistability branch, while Sec.~\ref{sec:Passive_Lower} is devoted to the study of the physics in the lower bistability branch.

Even though our predictions are fully general, in the following we focus on parameters inspired from specific integrated photonics devices. In detail, we consider that the coherent pump drives the $\ell_{\rm P}=133$ mode.
We set the value of the radiative losses due to the bus waveguides equal to the non-radiative ones, giving a quality factor of $Q\simeq 4.2\times 10^{5}$ (corresponding to $k_{\rm w}=k_{\rm A}=0.032$). As we are considering a ring resonator, no S waveguide is present and therefore $\gamma_{\rm S}=0$. We take an incident pump intensity $|\tilde{a}^{\rm (in)}_{\rm P+}|^2=1$ W, a nonlinear refractive index $n_{\rm NL}=10^{-14}$ cm$^2/$W, and we consider a device featuring waveguides of transverse area $A\simeq 0.70$ $\mu$m$^2$.

%
%%%%%%%%%%%%%%%%%%%%%%%%%%%%%%%%%%%%%%%%%%%%%%%%%%%%
\subsection{Upper bistability branch}
\label{sec:Passive_Upper}
%%%%%%%%%%%%%%%%%%%%%%%%%%%%%%%%%%%%%%%%%%%%%%%%%%%%
%

%
\begin{figure*}
    \centering
    \includegraphics[width=\textwidth]{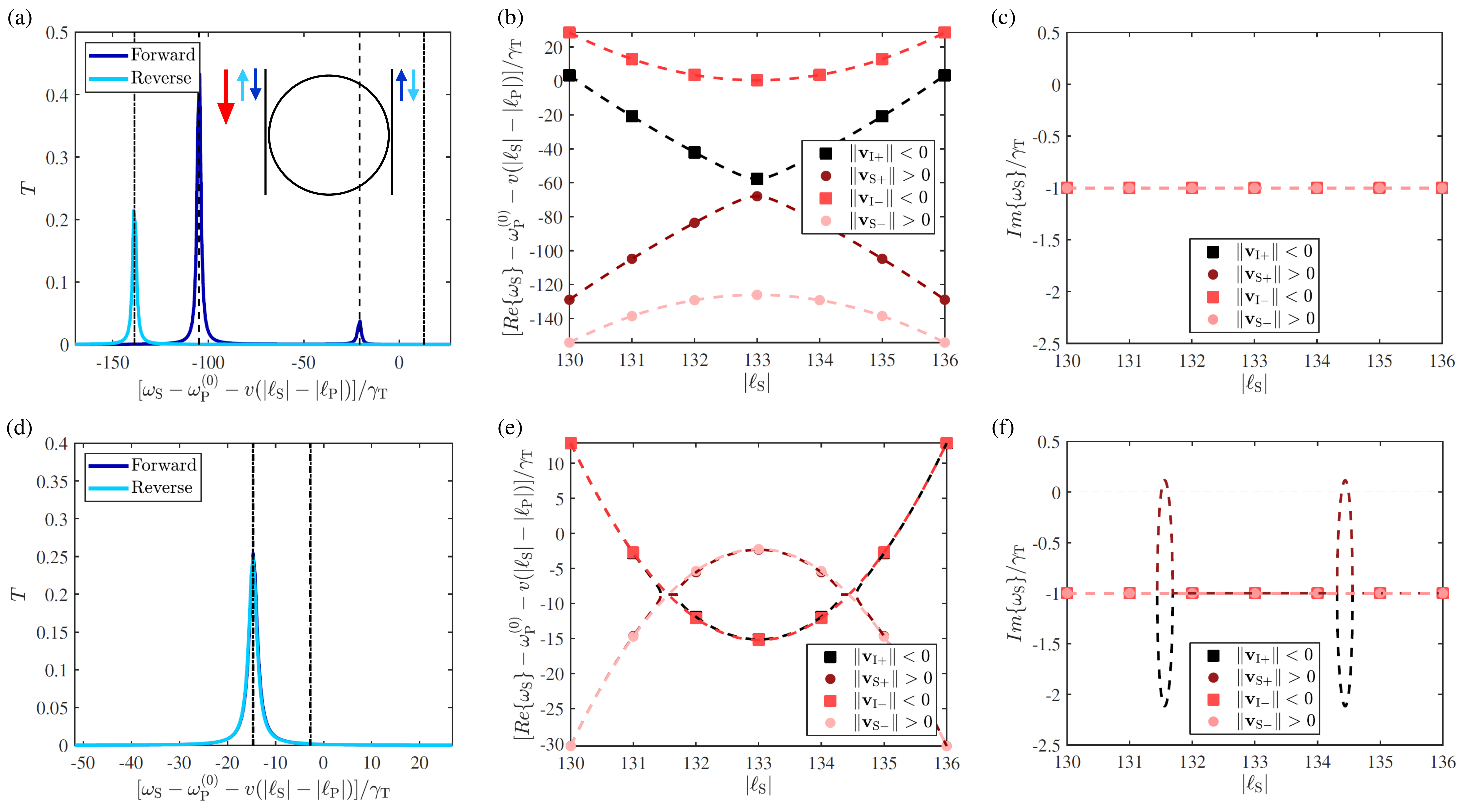}
    \caption{\textbf{(a,d)} Transmittance $T$ of a weak signal beam in the forward (dark blue) and reverse (cyan) directions across a coherently illuminated passive ring resonator in add-drop configuration as a function of the signal frequency $\omega_{\rm S}$ in the vicinity of the resonator modes with absolute angular momenta $|\ell_{\rm S}|=131$. The pump is close to resonance with the mode of angular momentum $\ell_{\rm P}=133$. Panels (a) and (d) display the results when we pump in the upper and lower bistability branches, respectively.   The vertical dashed-dotted and dashed lines signal respectively the real parts of the Bogoliubov eigenvalues corresponding to the reverse and forward-propagating signals. A sketch of the device is found in the upper-right part of panel (a). The big, red arrow represents the incident pump, while the small arrows describe the probing with the signal in the forward (dark blue) and in the reverse (cyan) directions. \textbf{(b,c)} Real and imaginary parts of the Bogoliubov eigenvalues $\omega_{\rm S}$ for a pump in the upper bistability branch as a function of the signal angular momenta $\pm|\ell_{\rm S}|$. The eigenvalues are labeled according to the Bogoliubov norm $\|\mathbf{v}_{\rm S,I\pm}\|$ of the associated eigenvectors. Squares and circles are the values corresponding to the modes of integer angular momenta $|\ell_{\rm S}|$. The dashed lines are calculated for non-integer values of $|\ell_{\rm S}|$. Panels \textbf{(e,f)} show analogous plots when the pump is on the lower bistability branch. The horizontal dashed pink line in panel (f) indicates the position of zero.}
    \label{fig:PassiveDiode}
\end{figure*}

Let us first assume that the system is prepared on the upper branch of the bistability loop shown in Fig.~\ref{fig:DiodeBistability}. Such a state can be prepared, e.g., by means of a downwards ramp of the pump frequency~\cite{MunozDeLasHeras_2020}. In this regime, the large intensity in the pump mode enhances the effect of FWM. To maximize this effect, we place the pump at a frequency $\omega_{\rm P}-\omega^{(0)}_{\rm P}\simeq -63\gamma_{\rm T}$, close to the downwards intensity jump of the upper branch, which yields a resonant enhancement of the pump intensity inside the resonator $|\tilde{a}^{(0)}_{\rm P+}|^2\simeq 240|\tilde{a}^{\rm (in)}_{\rm P+}|^2$.

% In Sec.~\ref{sec:Passive_Upper_Transmittance}
% we assess the transmittance of small amplitude signals in the forward and reverse directions. In Sec.~\ref{sec:Passive_Upper_Bogoliubov} we set up the Bogoliubov system of equations for signal and idler and diagonalize the Bogoliubov matrix in order to gain insight into the role of nonreciprocal FWM processes on the transmittance spectrum.

%
%%%%%%%%%%%%%%%%%%%%%%%%%%%%%%%%%%%%%%%%%%%%%%%%%%%%
\subsubsection{Transmission spectra}
\label{sec:Passive_Upper_Transmittance}
%%%%%%%%%%%%%%%%%%%%%%%%%%%%%%%%%%%%%%%%%%%%%%%%%%%%
%

We start by probing the device in the {\em forward} direction by sending a signal of (weak) incident intensity $|\tilde{a}^{\rm (in)}_{\rm S+}|^2$ through the left-hand bus waveguide. The forward transmittance $T_{\rm for}$ is plotted in Fig.~\ref{fig:PassiveDiode}a as a function of the signal frequency $\omega_{\rm S}$. Even though we are exploring the neighborhood of a single $\ell_{\rm S}=131$ mode, the spectrum is characterized by a doublet of peaks: The left-hand peak is in the rough vicinity of the signal mode shifted by the Kerr nonlinear terms in Eq.~\eqref{eq:GeneralS}. The right-hand peak, on the other hand, appears as a consequence of FWM processes and its position is fixed by the conservation of energy and momentum in the pump-mediated coupling with the idler. 

We now send a {\em reverse}-propagating signal of identical incident intensity $|\tilde{a}^{\rm (in)}_{\rm S-}|^2=|\tilde{a}^{\rm (in)}_{\rm S+}|^2$ through the right-hand bus waveguide and we look at the transmittance $T_{\rm rev}$ as its frequency $\omega_{\rm S}$ is varied in the same spectral region around the $\ell_{\rm S}=-131$ mode (of course resonant with the $\ell_{\rm S}=131$ one at linear regime, since we are considering a non-magnetic device), plotted as a cyan line in Fig.~\ref{fig:PassiveDiode}a. Since FWM is absent in this case, no photons from the pump are scattered into the idler resonance, and therefore $T_{\rm rev}$ displays a single peak. For the same reason, the spectral position of this peak is located at a slightly different frequency from that of the left-hand peak of the $T_{\rm for}$ doublet and is completely determined by the resonance shift given by the Kerr nonlinearity.
%This is also due to the lack of FWM coupling between pump, signal, and idler: in the reverse direction the only contribution to the nonlinear shift of the resonance peak comes from the Kerr nonlinearity. As can be seen in Fig.~\ref{fig:PassiveDiode}a, the reverse transmittance $T_{\rm rev}$ is negligible over a broad frequency range, including the frequencies at which the doublet in $T_{\rm for}$ appears. In particular, at the frequency $[\omega_{\rm S}-\omega^{(0)}_{\rm P}-v(|\ell_{\rm S}|-|\ell_{\rm P}|)]=-103.47\gamma_{\rm T}$ in which the forward-propagating signal achieves its maximum transmittance, the ratio between the two transmittances reaches $33$ dB, thus suggesting an efficient breaking of reciprocal behaviour and a promising optical isolation behaviour.

To summarize, our calculations anticipate that a nonlinear ring resonator can be made non-reciprocal by illuminating it with a strong pump beam. Non-reciprocity then leads to an efficient optical isolation for weak signal beams tuned close to resonance with neighboring modes of the resonator. Since the underlying equations are linear in the forward and reverse signal amplitudes $\tilde{a}_{\rm S\pm}$ and $\tilde{a}_{\rm I\pm}$, the signals propagating in opposite directions do not interact with each other and therefore the transmittance spectra in the two directions are unaltered even when the two signals are simultaneously propagating though the device. As a consequence the resonator is not subjected to the dynamic reciprocity restrictions and hence it can work as an effective optical isolator.

%
%%%%%%%%%%%%%%%%%%%%%%%%%%%%%%%%%%%%%%%%%%%%%%%%%%%%
\subsubsection{Bogoliubov analysis}
\label{sec:Passive_Upper_Bogoliubov}
%%%%%%%%%%%%%%%%%%%%%%%%%%%%%%%%%%%%%%%%%%%%%%%%%%%%
%

The nonreciprocal effect of FWM for counterpropagating signals can be further investigated by means of a linearized analysis of the dynamics of signal and idler fields in the forward and reverse directions. Here, the fields are treated within linear response theory as weak perturbations on top of the large intensity pump field that is present in the forward direction only. From Eqs.~(\ref{eq:GeneralS}-\ref{eq:GeneralI}) it is straightforward to set up the Bogoliubov system of equations
\begin{widetext}
\begin{align}
\scriptsize
i & \frac{d}{dt}
\begin{bmatrix}
\tilde{a}_{\rm S+} &
\tilde{a}^{*}_{\rm I+} &
\tilde{a}_{\rm S-} &
\tilde{a}^{*}_{\rm I-}
\end{bmatrix}^{T}
\nonumber\\
& =
\begin{bmatrix}
%(1,1)
\begin{matrix}
\omega^{(0)}_{\rm S}-\omega_{\rm S}-i\gamma_{\rm T}\\
-2g_{\rm NL}|\tilde{a}^{(0)}_{\rm P+}|^2
\end{matrix}
&
%(1,2)
\begin{matrix}
-g_{\rm NL}\tilde{a}^{(0)^2}_{\rm P+}
\end{matrix}
&
%(1,3)
\begin{matrix}
0
\end{matrix}
&
%(1,4)
\begin{matrix}
0
\end{matrix}
\\
%(2,1)
\begin{matrix}
+g_{\rm NL}\tilde{a}^{(0)^{*2}}_{\rm P+}
\end{matrix}
&
%(2,2)
\begin{matrix}
-\omega^{(0)}_{\rm I}+2\omega_{\rm P}-\omega_{\rm S}-i\gamma_{\rm T}\\
+2g_{\rm NL}|\tilde{a}^{(0)}_{\rm P+}|^2
\end{matrix}
&
%(2,3)
\begin{matrix}
0
\end{matrix}
&
%(3,4)
\begin{matrix}
0
\end{matrix}
\\
%(3,1)
\begin{matrix}
0
\end{matrix}
&
%(3,2)
\begin{matrix}
0
\end{matrix}
&
%(3,3)
\begin{matrix}
\omega^{(0)}_{\rm S}-\omega_{\rm S}-i\gamma_{\rm T}\\
-2g_{\rm NL}|\tilde{a}^{(0)}_{\rm P+}|^2
\end{matrix}
&
%(3,4)
\begin{matrix}
0
\end{matrix}
\\
%(4,1)
\begin{matrix}
0
\end{matrix}
&
%(4,2)
\begin{matrix}
0
\end{matrix}
&
%(4,3)
\begin{matrix}
0
\end{matrix}
&
%(4,4)
\begin{matrix}
-\omega^{(0)}_{\rm I}+2\omega_{\rm P}-\omega_{\rm S}-i\gamma_{\rm T}\\
+2g_{\rm NL}|\tilde{a}^{(0)}_{\rm P+}|^2
\end{matrix}
\end{bmatrix}
\nonumber\\
& \times
\begin{bmatrix}
\tilde{a}_{\rm S+} &
\tilde{a}^{*}_{\rm I+} &
\tilde{a}_{\rm S-} &
\tilde{a}^{*}_{\rm I-}
\end{bmatrix}^{T}
.
\label{eq:BogoMatrixPassiveRing}
\end{align}
\end{widetext}
The next step is to diagonalize the $4 \times 4$ matrix above in order to find the eigenvalues $\omega_{\rm S}$ and their corresponding eigenvectors $\mathbf{v}$ for several pairs of counterpropagating signal modes of opposite angular momentum $\pm|\ell_{\rm S}|$ in a range of $|\ell_{\rm S}|$ centered around $\ell_{\rm P}$. As usual, the real part of the eigenvalues $Re\{\omega_{\rm S}\}$ will tell us the position of the transmittance peaks for each pair of signals. The imaginary part, instead, gives information on the linewidth and the dynamical stability/instability of the configuration: a positive imaginary part $Im\{\omega_{\rm S}\}>0$ for some $\ell_{\rm S}$ implies that the system is dynamically unstable around that mode. %A negative imaginary part $Im\{\omega_{\rm S}\}<0$ at $\ell_{\rm S}$ signals instead the dynamical stability of that mode.

The four eigenvectors $\mathbf{v}=[v_1, v_2, v_3, v_4]^{T}$ can be classified according to their Bogoliubov norm,%
\begin{align}
    \|\mathbf{v}\|^2 = |v_1|^2-|v_2|^2+|v_3|^2-|v_4|^2,
\end{align}
which quantifies the coupling between signal and idler. Choosing for each eigenvector the normalization
\begin{align}
    {v^2_1+v^2_2+v^2_3+v^2_4} = 1\,,
\label{eq:NormEigenvectors}
\end{align}
it follows from Eq.~\eqref{eq:NormEigenvectors} that the Bogoliubov norm is restricted to values in the range
\begin{align}
    -1\leq\|\mathbf{v}\|\leq 1\,:
\end{align}
a negative norm implies an idler-dominated response, while a positive norm corresponds to a signal-dominated one.
Diagonalization of the matrix in Eq.~\eqref{eq:BogoMatrixPassiveRing} for the specific set of parameters considered in Fig.~\ref{fig:PassiveDiode}a gives four eigenvectors that can be labeled as $\mathbf{v}_{\rm I+}$, $\mathbf{v}_{\rm S+}$, $\mathbf{v}_{\rm I-}$, and $\mathbf{v}_{\rm S-}$ according to their respective signal/idler content, 
\begin{align}
    &\|\mathbf{v}_{\rm I+}\|=-0.724,\;\;
    \|\mathbf{v}_{\rm S+}\|=0.724,\;\;
    \nonumber\\
    &\|\mathbf{v}_{\rm I-}\|=-1,\;\;
    \|\mathbf{v}_{\rm S-}\|=1.
\end{align}
As expected, the norms of the eigenvectors belonging to the reverse-propagating idler $\mathbf{v}_{\rm I-}$ and signal $\mathbf{v}_{\rm S-}$ take the extreme values $\|\mathbf{v}_{\rm I-}\|=-1$ and $\|\mathbf{v}_{\rm S-}\|=1$, signaling that there is no coupling between them and each of them has a purely idler or signal nature. On the other hand, the norms of the eigenvectors belonging to the forward-propagating idler $\mathbf{v}_{\rm I+}$ and signal $\mathbf{v}_{\rm S+}$ are pushed closer to zero by the FWM terms that mix the signal/idler characters.

%We will label the eigenvalues associated to these eigenvectors as $\omega_{\rm S}(\mathbf{v}_{\rm I+})$, $\omega_{\rm S}(\mathbf{v}_{\rm S+})$, $\omega_{\rm S}(\mathbf{v}_{\rm I-})$, and $\omega_{\rm S}(\mathbf{v}_{\rm S-})$, respectively.
Panels (b) and (c) of Fig.~\ref{fig:PassiveDiode} display the real and imaginary parts of the four eigenvalues as a function of the absolute value of the angular momentum $|\ell_{\rm S}|$ of the pair of counterpropagating signals. 
To facilitate reading, in these panels we have added the results for non-integer values of $|\ell_{\rm S}|$. Although they are not relevant to study the transmittance across our resonator, since only integer values of $|\ell_{\rm S}|$ are physically meaningful and accessible, such intermediate regions allow us to understand the physics with more clarity.

For the reverse-propagating signal, it is evident from the form of the Bogoliubov matrix in Eq.~\eqref{eq:BogoMatrixPassiveRing} that the real part of the eigenvalue $\omega_{\rm S}(\mathbf{v}_{\rm S-})$ is solely determined by the Kerr nonlinearity shift of the resonance frequency of the signal mode with angular momentum $-\ell_{\rm S}$, 
\begin{align}
    Re\{\omega_{\rm S}(\mathbf{v}_{\rm S-})\}=\omega^{(0)}_{\rm S}-2g_{\rm NL}|\tilde{a}^{(0)}_{\rm P+}|^2.
\label{eq:ReOmega3}
\end{align}
On the other hand, the real part of the eigenvalue $\omega_{\rm S}(\mathbf{v}_{\rm I-})$ is given by the frequency at which FWM would couple the pump light into the signal mode if the FWM processes into the idler were active. Also this resonance is shifted by the Kerr nonlinearity, yet in the opposite direction,
\begin{align}
    Re\{\omega_{\rm S}(\mathbf{v}_{\rm I-})\}=2\omega_{\rm P}-\omega^{(0)}_{\rm I}+2g_{\rm NL}|\tilde{a}^{(0)}_{\rm P+}|^2.
\label{eq:ReOmega4}
\end{align}

As shown in Fig.~\ref{fig:PassiveDiode}b the real parts of these eigenvalues form two branches separated by a gap whose frequency width is minimum at the angular momentum of the pump $\ell_{\rm P}$. To highlight that these frequencies indeed correspond to the maxima of the reverse transmittance $T_{\rm rev}$, we plotted in Fig.~\ref{fig:PassiveDiode}a a dashed-dotted line at the frequency $Re\{\omega_{\rm S}(\mathbf{v}_{\rm S-})\}$ corresponding to the mode with angular momentum $\ell_{\rm S}=-131$. We also plotted a second dashed-dotted line at the frequency $Re\{\omega_{\rm S}(\mathbf{v}_{\rm I-})\}$ belonging to the mode $\ell_{\rm S}=-131$. 
The fact that no peak is displayed by $T_{\rm rev}$ at this position can be readily explained by the Bogoliubov norm of its associated eigenvector $\|\mathbf{v}_{\rm I-}\|=-1$: since this eigenvalue only features a contribution from the idler and FWM is not present for the reverse signal, the corresponding peak has a vanishing weight in the signal transmittance. Regarding the imaginary parts, both of them are given by the total loss rate of the resonator, i.e. $Im\{\omega_{\rm S}(\mathbf{v}_{\rm S,I-})\}=-\gamma_{\rm T}$ for all values of angular momentum $|\ell_{\rm S}|$ (Fig.~\ref{fig:PassiveDiode}c), which implies dynamical stability of our configuration against reverse-propagating perturbations for all values of $\ell_{\rm S}$.

On the other hand, the eigenvalues $\omega_{\rm S}(\mathbf{v}_{\rm S,I+})$ associated to forward-propagating signals are given by the diagonalization of the top-left $2 \times 2$ block of the Bogoliubov matrix. Again, their real parts give rise to a couple of branches separated by a gap, as shown in Fig.~\ref{fig:PassiveDiode}b. However, in this case, the presence of FWM results in an effective attraction between the two branches. The frequency gap between them is therefore reduced but remains nevertheless open, with a minimum frequency width at the pump angular momentum $\ell_{\rm P}$. To highlight that $Re\{\omega_{\rm S}(\mathbf{v}_{\rm S,I+})\}$ indeed correspond to the peaks in $T_{\rm for}$, their values at an angular momentum  $\ell_{\rm S}=131$ are plotted as dashed vertical lines in Fig.~\ref{fig:PassiveDiode}a: here, both real parts agree perfectly with the transmittance maxima. The imaginary parts, plotted in Fig.~\ref{fig:PassiveDiode}c, form a flat band at a negative value given by the total loss rate $Im\{\omega_{\rm S}(\mathbf{v}_{\rm S,I+})\}=-\gamma_{\rm T}$ independently of $|\ell_{\rm S}|$, which implies that the pump is dynamically stable also against forward-propagating perturbations.

%
%%%%%%%%%%%%%%%%%%%%%%%%%%%%%%%%%%%%%%%%%%%%%%%%%%%%
\subsubsection{Optical isolation on the pump mode}
\label{sec:Passive_Pump_Mode}
%%%%%%%%%%%%%%%%%%%%%%%%%%%%%%%%%%%%%%%%%%%%%%%%%%%%
%

%
\begin{figure}
    \includegraphics[width=0.5\textwidth]{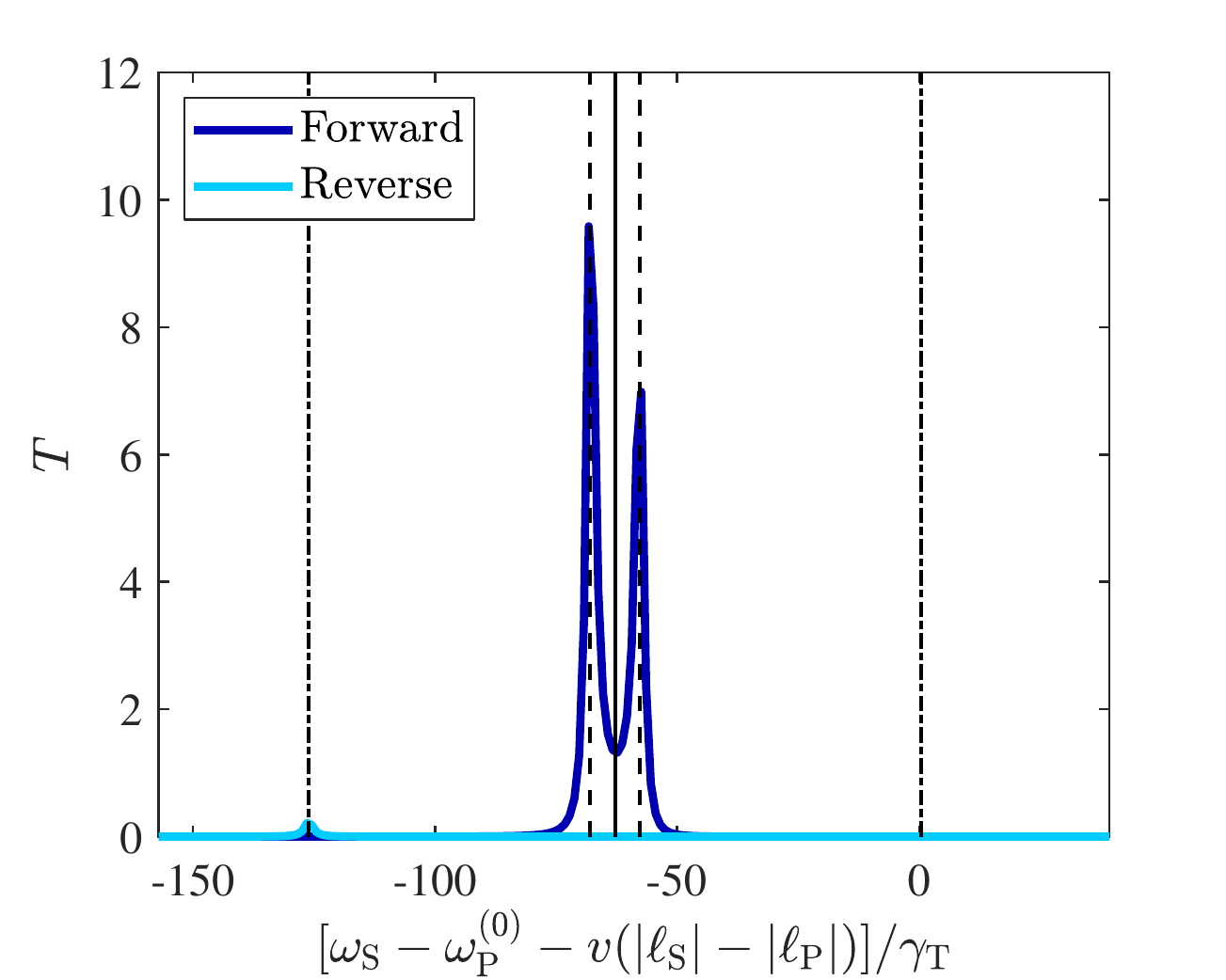}
    \caption{Transmittance $T$ in forward (dark blue) and reverse (cyan) directions through an optically pumped passive ring resonator in add-drop configuration as a function of the signal frequency $\omega_{\rm S}$ for forward- or reverse-propagating signal modes with absolute angular momenta equal to the pump one, $|\ell_{\rm S}|=\ell_{\rm P}=133$. The vertical dashed-dotted and dashed lines signal respectively the real parts of the Bogoliubov eigenvalues corresponding to the reverse and forward-propagating signals. The vertical solid line indicates the pump frequency $\omega_{\rm P}$. Same pump parameters as in Fig.~\ref{fig:PassiveDiode}a-c.}
    \label{fig:Passive_PumpMode}
\end{figure}

As a last point, we look in closer detail at the transmittance of small-amplitude signals whose frequency is varied around the resonance of the pump mode, i.e. signals featuring angular momenta $|\ell_{\rm S}|=\ell_{\rm P}=133$. This case is of special interest as it allows a direct comparison with the reasoning of Ref.~\cite{Shi_2015}. %, where the spatial dependence of the electric field was ruled out.

Fig.~\ref{fig:Passive_PumpMode} shows our results for the transmittance of forward and reverse-propagating signals for a pump on the upper bistability branch. As in the previous case, transmittance is remarkably different in the two directions: in the forward one (dark blue line) the usual doublet is visible around the red-detuned pump frequency $\omega_{\rm P}$ (indicated by the vertical solid line). On the other hand, the transmittance in the reverse direction (cyan line) displays a single peak given by the Kerr nonlinearity shift, as accounted by Eq.~\eqref{eq:ReOmega3}. The non-reciprocal behavior is further enhanced in the vicinity of the pump frequency by FWM-induced amplification of the forward transmitted signal at a frequency position where the reverse transmittance is instead very small. For the parameters in the figure, the transmittance difference in the two directions reaches a value of up to $52$ dB at the frequency $[\omega_{\rm S}-\omega^{(0)}_{\rm P}-v(|\ell_{\rm S}|-|\ell_{\rm P}|)]=-67.98\gamma_{\rm T}$ of the left component of the doublet.

This observation generalizes the result of~\cite{DelBino_2018} for monochromatic light to a finite range of frequencies surrounding the pump one: interestingly, even though we are considering pump and signal as monochromatic waves whose frequency spectra do not overlap, the asymmetry in the FWM coupling leads to a sizable nonreciprocal transmittance. Our results therefore expand the range of situations in which the dynamic reciprocity introduced in Ref.~\cite{Shi_2015} does not apply.

%
%%%%%%%%%%%%%%%%%%%%%%%%%%%%%%%%%%%%%%%%%%%%%%%%%%%%
\subsection{Lower bistability branch}
\label{sec:Passive_Lower}
%%%%%%%%%%%%%%%%%%%%%%%%%%%%%%%%%%%%%%%%%%%%%%%%%%%%
%

%Note that no unstable behaviors arise in the upper branch of the optical bistability as the imaginary parts of the eigenvalues are always negative. As we mentioned before this is 

While the upper bistability branch considered before is the most promising choice for optical isolation as it provides a larger pump intensity that enhances FWM, for the sake of completeness it is interesting to have a look also at the behavior on the lower bistability branch. To this purpose, we set the pump frequency to $\omega_{\rm P}-\omega^{(0)}_{\rm P}\simeq -8.7\gamma_{\rm T}$ close to the upwards intensity jump shown in Fig.~\ref{fig:DiodeBistability} and we assume an appropriate upwards frequency ramp to obtain a pump intensity $|\tilde{a}^{(0)}_{\rm P+}|^2\simeq 4.3|\tilde{a}^{\rm (in)}_{\rm P}|^2$ on the lower branch.

%We start by looking at the transmittance spectrum in Sec.~\ref{sec:Passive_Lower_Transmittance}. We then employ a Bogoliubov analysis in Sec.~\ref{sec:Passive_Lower_Bogoliubov}.

% %
% %%%%%%%%%%%%%%%%%%%%%%%%%%%%%%%%%%%%%%%%%%%%%%%%%%%%
% \subsubsection{Transmittance}
% \label{sec:Passive_Lower_Transmittance}
% %%%%%%%%%%%%%%%%%%%%%%%%%%%%%%%%%%%%%%%%%%%%%%%%%%%%
% %

As it is shown in Fig.~\ref{fig:PassiveDiode}d, in this case the pump intensity is not sufficiently large to give rise to an appreciable FWM. As a consequence, the transmittance spectrum $T_{\rm for,rev}(\omega_{\rm S})$ for the pair of counterpropagating modes with absolute angular momenta $|\ell_{\rm S}|=131$ is practically identical in the two directions and features a single peak at a frequency $\omega_{\rm S}=\omega^{(0)}_{\rm S}-2g_{\rm NL}|\tilde{a}^{(0)}_{\rm P+}|^2$ determined by the Kerr nonlinearity shift of the resonance frequency.

% %
% %%%%%%%%%%%%%%%%%%%%%%%%%%%%%%%%%%%%%%%%%%%%%%%%%%%%
% \subsubsection{Bogoliubov analysis}
% \label{sec:Passive_Lower_Bogoliubov}
% %%%%%%%%%%%%%%%%%%%%%%%%%%%%%%%%%%%%%%%%%%%%%%%%%%%%
% %

Similarly to what we did for the upper bistability branch, we diagonalize the Bogoliubov matrix in Eq.~\eqref{eq:BogoMatrixPassiveRing} and obtain the eigenvalues and eigenvectors. In this case the Bogoliubov norms of the four eigenvectors are
\begin{align}
    &\|\mathbf{v}_{\rm I+}\|=-0.999,\;\;
    \|\mathbf{v}_{\rm S+}\|=0.999,\;\;
    \nonumber\\
    &\|\mathbf{v}_{\rm I-}\|=-1,\;\;
    \|\mathbf{v}_{\rm S-}\|=1,
\end{align}
signaling a finite but practically negligible coupling between signal and idler in the forward case. So similar values of the norms in the forward and reverse signals explain the almost identical transmittance in the two directions.

The real and imaginary parts of the Bogoliubov eigenvalues $\omega_{\rm S}(\mathbf{v}_{\rm S,I\pm})$ are shown in Fig.~\ref{fig:PassiveDiode}e,f, respectively. The two real parts $Re\{\omega_{\rm S}(\mathbf{v}_{\rm S,I-})\}$ belonging to the reverse-propagating signal intersect at two non-integer values of angular momentum. However, given the absence of FWM processes, their imaginary parts are still given by the total loss rate $Im\{\omega_{\rm S}(\mathbf{v}_{\rm S,I-})\}=-\gamma_{\rm T}$ as in the upper bistability branch. %This grants the dynamical stability of the pump against reverse signals also in the lower bistability branch. 
The situation is more interesting in the case of the forward-propagating signal: for non-integer angular momenta around the two intersection points of $Re\{\omega_{\rm S}(\mathbf{v}_{\rm S,I-})\}$ the two real parts $Re\{\omega_{\rm S}(\mathbf{v}_{\rm S,I+})\}$ coalesce and the corresponding imaginary parts $Im\{\omega_{\rm S}(\mathbf{v}_{\rm S,I+})\}$ are split. When the pump frequency $\omega_{\rm P}$ approaches the upwards intensity jump of the bistability loop, the imaginary part of $\omega_{\rm S}(\mathbf{v}_{\rm S+})$ turns positive in the aforementioned regions of $\ell_{\rm S}$. However, since this occurs at unphysical non-integer values of angular momentum, the anticipated instability is immaterial and has no practical consequence for the chosen parameters.

A physical instability could nevertheless appear if the pump parameters were slightly modified so to shift the instability to integer values of angular momentum and/or by extending the width of the instability region in a device with a smaller (negative) curvature $\alpha$ of the dispersion relation. If some integer values of $\ell_{\rm S}$ fall within the instability region, the system turns dynamically unstable towards optical parametric oscillation into the forward-propagating signal and idler modes.

%
%%%%%%%%%%%%%%%%%%%%%%%%%%%%%%%%%%%%%%%%%%%%%%%%%%%%
\subsection{Connection with quantum fluids of light}
%%%%%%%%%%%%%%%%%%%%%%%%%%%%%%%%%%%%%%%%%%%%%%%%%%%%
%

As a final remark, it is interesting to note how the physics presented in this Section for strongly pumped nonlinear ring resonators in integrated photonics can be connected to the behaviour of quantum fluids of light in planar microcavities under the so-called coherent pumping~\cite{Carusotto_2013}. In particular, very similar transmission spectra arise when one studies the elementary excitations around the steady-state. Also in planar cavity devices, the photon density features a bistable behavior, the collective excitations arise from FWM processes between the different bands, and can be probed by weak additional signal beams, as experimentally demonstrated in~\cite{claude2021highresolution}. In fluids of light, a non-reciprocal transmission of a weak signal beam naturally appears as soon as the coherently-pumped fluid of light is moving at a constant velocity $v$, e.g. when it is generated by a finite-$k$ coherent pump.

In more formal terms, it is instructive to compare Eq.~\eqref{eq:DispersionRelationLambda}, which accounts for the dispersion of photon modes in our ring resonator, with the dispersion of photons along the planar microcavity. As it is spelled out in detail in Appendix~\ref{sec:Appendix}, the bare resonance frequency $\omega^{(0)}_{\rm P}$ of the $\ell_{\rm P}$ mode is analogous to the bare frequency of the pumped mode, the linear slope of the dispersion $v(|\ell|-|\ell_{\rm P}|)$ plays the role of the group velocity of the mode, the curvature $\alpha(|\ell|-|\ell_{\rm P}|)^2/2$ is related to the inverse photon mass, and the Kerr nonlinearity plays the role of the photon-photon interactions. 

All together, one can then think of the optical field in the ring resonator as a fluid of light that circulates around the resonator at angular speed $v$: the marked attraction between the S+ and the I+ bands visible in Fig.~\ref{fig:PassiveDiode}b is a precursor of a sonic dispersion, the appearance of the I+ peak in Fig.~\ref{fig:PassiveDiode}a corresponds to the ghost branch observed in~\cite{claude2021highresolution}, and the optical isolation behaviour induced by the FWM can be interpreted as stemming from the dragging of collective excitations by the moving fluid. As we are going to see in the next Sections, this deep analogy provides a useful guidance also in the case of incoherently pumped lasing resonators.

%%%%%%%%%%%%%%%%%%%%%%%%%%%%%%%%%%% PUMPED ACTIVE RING RESONATOR
%%%%%%%%%%%%%%%%%%%%%%%%%%%%%%%%%%

%
\begin{figure*}
    \centering
    \includegraphics[width=\textwidth]{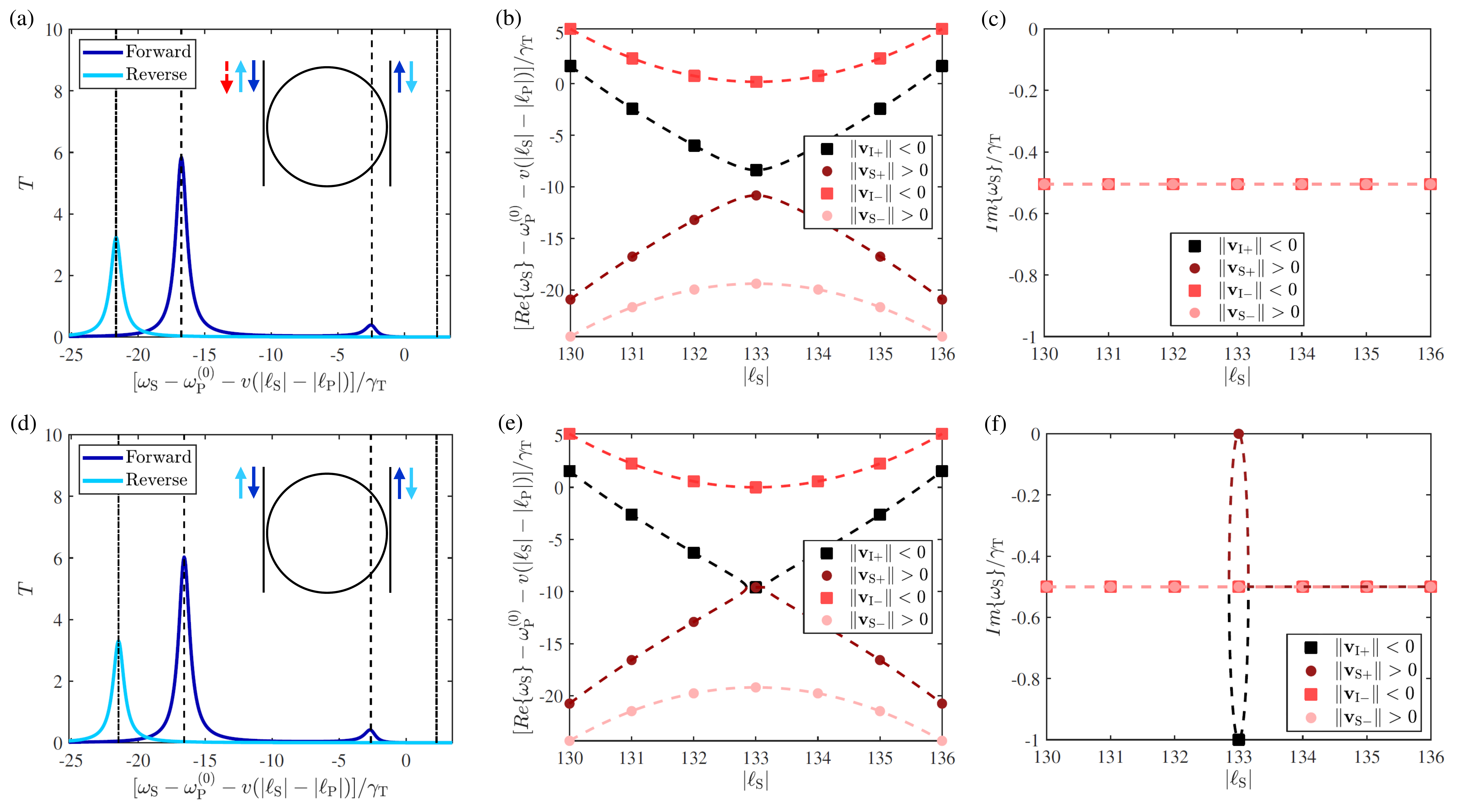}
    \caption{\textbf{(a,d)} Transmittance $T$ in forward (dark blue) and reverse (cyan) directions across an incoherently pumped active ring resonator in add-drop configuration lasing in the $\ell_{\rm P}=133$ mode, as a function of the signal frequency $\omega_{\rm S}$ for a pair of counterpropagating modes with angular momenta $|\ell_{\rm S}|=131$. 
    The pump rate is fixed at $P_{0}=2\gamma_{\rm T}$. A sketch of the device is found in the upper-right part of panel (a). The red, dashed arrow represents the weak incident seed, while the solid arrows describe the signals probing the device. The vertical dashed-dotted and dashed lines indicate the reals part of the Bogoliubov eigenvalues corresponding to the reverse and forward-propagating signals. \textbf{(b,c)} and \textbf{(e,f)} Real and imaginary parts of the Bogoliubov eigenvalues $\omega_{\rm S}$ as the angular momenta $\pm|\ell_{\rm S}|$ of two signals propagating in the forward and reverse directions are varied around $\pm\ell_{\rm P}=\pm 133$. The eigenvalues are labeled according to the Bogoliubov norm $\|\mathbf{v}_{\rm S,I\pm}\|$ of the corresponding eigenvectors. Squares and circles indicate the modes of integer angular momenta $|\ell_{\rm S}|$, while the dashed lines are calculated for non-integer values of $|\ell_{\rm S}|$. Panels \textbf{(a-c)} and \textbf{(d-f)} display the results in the presence/absence of a weak seed deterministically triggering lasing in the CCW direction, respectively. 
    }
\label{fig:RingLaserDiode}
\end{figure*}

\section{Active ring resonator}
\label{sec:TaijiDiode_ActiveRing}

While the coherent pump configuration investigated in the previous Section provides an efficient and controllable way to break reciprocity and induce an optical isolation behaviour, it has the crucial disadvantage of requiring an external coherent laser to pump the resonator. In this Section, we investigate a different configuration that alleviates this requirement and may provide a standalone isolating device.

Specifically, we investigate an active analog of the passive nonlinear ring resonator studied in the previous Section. By incoherently pumping the resonator above the lasing threshold $P_{0}>\gamma_{\rm T}$ we can induce lasing in one of the $\pm\ell_{\rm P}$ counterpropagating modes in which we assume gain to be the strongest. If backscattering is negligible, a local saturable gain will grant unidirectional lasing with a well-defined chirality in one of these two modes, as shown in Ref.~\cite{MunozDeLasHeras_2021b}. However, for the system to work as an isolator, we need to deterministically choose one of the operation directions as the forward one, otherwise the lasing chirality and thus the forward and reverse directions would be selected in a random way each time laser operation is switched on. Without loss of generality, we assume that the CCW direction is imposed as the forward one by means of some weak seed beam $\tilde{a}^{\rm (in)}_{\rm P+}$ coupling into the $\ell_{\rm P}>0$ mode and triggering laser operation in the CCW-propagating $\ell_{\rm P}$ mode. Once lasing in the desired direction is triggered, the seed can be of course switched off. 
%. Of course, changing the driving direction will result in the resonator lasing in the opposite mode with angular momentum $-\ell_{\rm P}$. Nevertheless, once lasing in the desired direction is triggered, one could switch off the driving. 
In what follows, we will first analyze in Sec.~\ref{sec:ActiveRing_Forced} the case where isolation is probed while the seed is still on, then in Sec.~\ref{sec:ActiveRing_Unforced} we will consider the case where the seed is turned off prior to probing optical isolation.

While our conclusions are fully general, for illustrative purposes we show simulations with reasonable parameters for integrated photonics devices, namely a ring-bus waveguide coupling $k_{\rm w}=0.04$, absorption losses $k_{\rm A}=0.032$ (yielding a quality factor $Q\simeq 7.6\times 10^{4}$ for the unloaded cavity), and an effective nonlinear coefficient $g_{\rm NL}n_{\rm S}/\gamma_{\rm T}\simeq 9.6$. As done for the passive ring resonator, we choose a pump mode of angular momentum $\ell_{\rm P}=133$. We set a pump rate $P_{0}=2\gamma_{\rm T}$, which gives a pump intensity inside the resonator $|\tilde{a}^{(0)}_{\rm P}|^2/n_{\rm S}=P_{0}/\gamma_{\rm T}-1=1$. The lasing frequency is determined by the Kerr nonlinear shift of the pump resonance frequency, i.e. $\omega_{\rm P}=\omega^{(0)}_{\rm P}-g_{\rm NL}n_{\rm S}$. 
%For each case, we start by looking at non-reciprocal features in the transmittance spectrum of small amplitude signals propagating in the forward and reverse directions, and then we interpret the result in terms of the Bogoliubov theory.

%
%%%%%%%%%%%%%%%%%%%%%%%%%%%%%%%%%%%%%%%%%%%%%%%%%%%%
\subsection{Optical isolation in the presence of a weak seed}
\label{sec:ActiveRing_Forced}
%%%%%%%%%%%%%%%%%%%%%%%%%%%%%%%%%%%%%%%%%%%%%%%%%%%%
%

In this Subsection we consider that a weak seed beam of amplitude $\tilde{a}^{\rm (in)}_{\rm P+}$ is driving the pump mode to deterministically enforce lasing into the CCW $\ell_{\rm P}=133$ mode. Specifically, we consider a seed beam with a weak incident intensity  $|\tilde{a}^{\rm (in)}_{\rm P+}|^2=10^{-2}n_{\rm S}$ at a frequency $\omega_{\rm P}$ perfectly matched with the natural laser oscillation at  $\omega^{(0)}_{\rm P}-g_{\rm NL}n_{\rm S}$. Similar behaviour would be found for slightly different seed frequencies, provided they remain in the vicinity of the natural lasing frequency. 

%
%%%%%%%%%%%%%%%%%%%%%%%%%%%%%%%%%%%%%%%%%%%%%%%%%%%%
\subsubsection{Transmission spectra}
\label{sec:ActiveRing_Forced_Transmittance}
%%%%%%%%%%%%%%%%%%%%%%%%%%%%%%%%%%%%%%%%%%%%%%%%%%%%
%

%In order to demonstrate isolation of reverse-propagating signals we follow the same strategy as in Sec.~\ref{sec:TaijiDiode_PassiveRing}. By means of a signal 

We probe optical isolation by comparing the transmittance of signal probes of weak amplitude $\tilde{a}^{(0)}_{\rm S\pm}$ propagating in either the forward or reverse directions, tuned in the vicinity of the resonance frequency $\omega^{(0)}_{\rm S}$ of the modes with angular momentum $|\ell_{\rm S}|=131$. The transmittance in the forward direction $T_{\rm for}$ is shown as a dark blue line in Fig.~\ref{fig:RingLaserDiode}a: as expected, the spectrum displays the characteristic doublet resulting from FWM. %We then probe the system using a reverse-propagating signal $\tilde{a}^{(0)}_{\rm S-}$ of the identical incident intensity $|\tilde{a}^{\rm (in)}_{\rm S-}|^2=|\tilde{a}^{\rm (in)}_{\rm S+}|^2$ that couples to the mode with opposite angular momentum $\ell_{\rm S}=-131$. 
The transmittance in the reverse direction $T_{\rm rev}$ is displayed as a cyan line in the same panel: due to the absence of FWM, the spectrum features a single  maximum at a frequency given by the Kerr nonlinearity shift of $\omega^{(0)}_{\rm S}$ and different from the ones of the FWM-induced doublet in $T_{\rm for}$. In particular, at the frequency $[\omega_{\rm S}-\omega^{(0)}_{\rm P}-v(|\ell_{\rm S}|-|\ell_{\rm P}|)]=-16.52\gamma_{\rm T}$ where $T_{\rm for}$ achieves its maximum value, the ratio between transmittance in the two directions reaches $23$ dB. As in the previous cases, the linearity of Eqs.~(\ref{eq:GeneralS}-\ref{eq:GeneralI}) with respect to the signal and idler fields makes the transmittance spectra $T_{\rm for,rev}$ to be unaffected by the presence of each other, therefore granting that our device can work as an efficient optical isolator.

Looking at the analytical form of the motion equations~(\ref{eq:GeneralS}-\ref{eq:GeneralI}), it is interesting to notice that FWM processes can also be induced by the saturable gain terms. This results in a finite FWM coupling between signal and idler even in active devices featuring negligible Kerr nonlinearities. However, the typical intensity of gain-induced FWM is always of the order of the saturable gain magnitude, and therefore of the loss rate $\gamma_{\rm T}$. On the other hand, the FWM stemming from the Kerr nonlinearity can be made arbitrarily larger with a suitable choice of the nonlinear material and increases with the pump rate. In particular, for the parameters employed in this Section we checked that at the optimal frequency mentioned above, %$\omega_{\rm S}=7.5132\times 10^{4} \gamma_{\rm T}$ 
the Kerr-originated FWM is about $20$ times larger than the gain-induced FWM. As a consequence, when the Kerr nonlinearity is removed by setting $g_{\rm NL}=0$, the transmittance spectrum displays an almost identical shape for signals propagating in the two directions with no isolation behavior.

%
%%%%%%%%%%%%%%%%%%%%%%%%%%%%%%%%%%%%%%%%%%%%%%%%%%%%
\subsubsection{Bogoliubov analysis}
\label{sec:ActiveRing_Forced_Bogoliubov}
%%%%%%%%%%%%%%%%%%%%%%%%%%%%%%%%%%%%%%%%%%%%%%%%%%%%
%

Like in the previous Section, the non-reciprocity underlying the different transmittance of the forward- and reverse-propagating signals can be physically understood in terms of the linearized equations of motion for the signal and idler fields in the two directions by diagonalizing the Bogoliubov matrix in a range of angular momenta $\pm|\ell_{\rm S}|$ around $\pm\ell_{\rm P}$. In the case of an active ring resonator the system of differential equations reads
\begin{widetext}
\begin{align}
i & \frac{d}{dt}
\begin{bmatrix}
\tilde{a}_{\rm S+} &
\tilde{a}^{*}_{\rm I+} &
\tilde{a}_{\rm S-} &
\tilde{a}^{*}_{\rm I-}
\end{bmatrix}^{T}
\nonumber\\
& =
\begin{bmatrix}
%(1,1)
\begin{matrix}
\omega^{(0)}_{\rm S}-\omega_{\rm S}-i\gamma_{\rm T}\\
-2g_{\rm NL}|\tilde{a}^{(0)}_{\rm P+}|^2
\\
+i\frac{P_{0}}{1+\frac{1}{n_{\rm S}}|\tilde{a}^{(0)}_{\rm P+}|^2}
\\
-i\frac{P_{0}/n_{\rm S}}{\left[1+\frac{1}{n_{\rm S}}|\tilde{a}^{(0)}_{\rm P+}|^2\right]^2}|\tilde{a}^{(0)}_{\rm P+}|^2
\end{matrix}
&
%(1,2)
\begin{matrix}
-g_{\rm NL}\tilde{a}^{(0)^2}_{\rm P+}
\\
-i\frac{P_{0}/n_{\rm S}}{\left[1+\frac{1}{n_{\rm S}}|\tilde{a}_{\rm P+}|^2\right]^2}\tilde{a}^{(0)^2}_{\rm P+}
\end{matrix}
&
%(1,3)
\begin{matrix}
0
\end{matrix}
&
%(1,4)
\begin{matrix}
0
\end{matrix}
\\
%(2,1)
\begin{matrix}
+g_{\rm NL}\tilde{a}^{(0)^{*2}}_{\rm P+}
\\
-i\frac{P_{0}/n_{\rm S}}{\left[1+\frac{1}{n_{\rm S}}|\tilde{a}^{(0)}_{\rm P+}|^2\right]^2}\tilde{a}^{(0)^{*2}}_{\rm P+}
\end{matrix}
&
%(2,2)
\begin{matrix}
-\omega^{(0)}_{\rm I}+2\omega_{\rm P}-\omega_{\rm S}-i\gamma_{\rm T}\\
+2g_{\rm NL}|\tilde{a}^{(0)}_{\rm P+}|^2
\\
+i\frac{P_{0}}{1+\frac{1}{n_{\rm S}}|\tilde{a}^{(0)}_{\rm P+}|^2}
\\
-i\frac{P_{0}/n_{\rm S}}{\left[1+\frac{1}{n_{\rm S}}|\tilde{a}^{(0)}_{\rm P+}|^2\right]^2}|\tilde{a}^{(0)}_{\rm P+}|^2
\end{matrix}
&
%(2,3)
\begin{matrix}
0
\end{matrix}
&
%(3,4)
\begin{matrix}
0
\end{matrix}
\\
%(3,1)
\begin{matrix}
0
\end{matrix}
&
%(3,2)
\begin{matrix}
0
\end{matrix}
&
%(3,3)
\begin{matrix}
\omega^{(0)}_{\rm S}-\omega_{\rm S}-i\gamma_{\rm T}\\
-2g_{\rm NL}|\tilde{a}^{(0)}_{\rm P+}|^2
\\
+i\frac{P_{0}}{1+\frac{1}{n_{\rm S}}|\tilde{a}^{(0)}_{\rm P+}|^2}\\
-i\frac{P_{0}/n_{\rm S}}{\left[1+\frac{1}{n_{\rm S}}|\tilde{a}^{(0)}_{\rm P+}|^2\right]^2}|\tilde{a}^{(0)}_{\rm P+}|^2
\end{matrix}
&
%(3,4)
\begin{matrix}
0
\end{matrix}
\\
%(4,1)
\begin{matrix}
0
\end{matrix}
&
%(4,2)
\begin{matrix}
0
\end{matrix}
&
%(4,3)
\begin{matrix}
0
\end{matrix}
&
%(4,4)
\begin{matrix}
-\omega^{(0)}_{\rm I}+2\omega_{\rm P}-\omega_{\rm S}-i\gamma_{\rm T}\\
+2g_{\rm NL}|\tilde{a}^{(0)}_{\rm P+}|^2
\\
+i\frac{P_{0}}{1+\frac{1}{n_{\rm S}}|\tilde{a}^{(0)}_{\rm P+}|^2}\\
-i\frac{P_{0}/n_{\rm S}}{\left[1+\frac{1}{n_{\rm S}}|\tilde{a}^{(0)}_{\rm P+}|^2\right]^2}|\tilde{a}^{(0)}_{\rm P+}|^2
\end{matrix}
\end{bmatrix}
\nonumber\\
& \times
\begin{bmatrix}
\tilde{a}_{\rm S+} &
\tilde{a}^{*}_{\rm I+} &
\tilde{a}_{\rm S-} &
\tilde{a}^{*}_{\rm I-}
\end{bmatrix}^{T}
.
\label{eq:BogoMatrixActive}
\end{align}
\end{widetext}
\normalsize
By diagonalizing the $4 \times 4$ matrix above we get the four eigenvectors $\mathbf{v}_{\rm S,I\pm}$ and their corresponding eigenvalues $\omega_{\rm S}(\mathbf{v}_{\rm S,I\pm})$, whose Bogoliubov norms 
\begin{align}
    &\|\mathbf{v}_{\rm I+}\|=-0.759,\;\;
    \|\mathbf{v}_{\rm S+}\|=0.759,\;\;
    \nonumber\\
    &\|\mathbf{v}_{\rm I-}\|=-1,\;\;
    \|\mathbf{v}_{\rm S-}\|=1,
\label{eq:BogoliubovNormActiveRing}
\end{align}
show a finite coupling between signal and idler only for the forward direction. 
The real and imaginary parts of the eigenvalues are displayed in Fig.~\ref{fig:RingLaserDiode}b,c as a function of the chosen mode. As usual, deeper insight about the underlying physics is obtained by plotting the whole spectrum of eigenvalues also for non-integer values of $\ell_{\rm S}$. Similarly to the passive ring resonator pumped in the upper branch of the bistability loop, the real parts are arranged in four branches separated by a gap that takes its minimum width at $|\ell_{\rm S}|=\ell_{\rm P}$. 

For the reverse-propagating signal, FWM is not present and the $Re\{\omega_{\rm S}(\mathbf{v}_{\rm S,I-})\}$ branches are completely determined by the Kerr nonlinear shift of the signal and idler modes, as determined by Eqs.~(\ref{eq:ReOmega3}-\ref{eq:ReOmega4}). The values of the two frequencies $Re\{\omega_{\rm S}(\mathbf{v}_{\rm S,I-})\}$ for $\ell_{\rm S}=-131$ are plotted as vertical dashed-dotted lines in Fig.~\ref{fig:RingLaserDiode}a: while $Re\{\omega_{\rm S}(\mathbf{v}_{\rm S-})\}$ can be associated to the maximum of $T_{\rm rev}$, no peak is found in $T_{\rm rev}$ at $Re\{\omega_{\rm S}(\mathbf{v}_{\rm I-})\}$: since FWM is not present for the reverse signal, no light from the pump can in fact be coupled at this frequency while simultaneously satisfying the conservation of energy and angular momentum.  The corresponding imaginary parts $Im\{\omega_{\rm S}(\mathbf{v}_{\rm S,I-})\}$ take the value
\begin{align}
\label{eq:Im_Forced}
    Im\{\omega_{\rm S}(\mathbf{v}_{\rm S,I-})\}=
    -\frac{P_0 - \gamma_{\rm T}}{P_0/\gamma_{\rm T}}
\end{align}
independently of $\ell_{\rm S}$. In our specific $P_0 = 2\gamma_{\rm T}$ case, we get $Im\{\omega_{\rm S}(\mathbf{v}_{\rm S,I-})\}=-\gamma_{\rm T}/2$ which confirms that the system is dynamically stable against perturbations propagating in the reverse direction.

As it is demonstrated by the Bogoliubov norms~\eqref{eq:BogoliubovNormActiveRing}, the situation is somehow different for the forward case. Here, FWM couples the pump with the resonance frequencies of signal and idler. The doublet of peaks in $T_{\rm for}$ corresponds to the two frequencies $Re\{\omega_{\rm S}(\mathbf{v}_{\rm S,I+})\}$ indicated as vertical dashed lines in Fig.~\ref{fig:RingLaserDiode}a: in agreement with the $\ell_{\rm S}$-dependent dispersion shown in Fig.~\ref{fig:RingLaserDiode}b, FWM results in an effective attraction between transmittance peaks and brings the two eigenvalues branches closer together narrowing the gap. Also in this case, the imaginary parts $Im\{\omega_{\rm S}(\mathbf{v}_{\rm S,I+})\}=-\gamma_{\rm T}/2$ form a flat band at the value in Eq.~\eqref{eq:Im_Forced}. This confirms the dynamical stability of the configuration also against perturbations propagating in the forward direction.

%
%%%%%%%%%%%%%%%%%%%%%%%%%%%%%%%%%%%%%%%%%%%%%%%%%%%%
\subsection{Optical isolation in the absence of the seed}
\label{sec:ActiveRing_Unforced}
%%%%%%%%%%%%%%%%%%%%%%%%%%%%%%%%%%%%%%%%%%%%%%%%%%%%
%

While the presence of a seed beam is useful to deterministically trigger lasing in the desired CCW direction in an otherwise symmetric ring resonator, there is no need to keep the seed on while probing the optical isolation behavior. In this Subsection, we show that the system performance remains indeed unchanged in the absence of the seed beam provided lasing is taking place in the desired CCW direction.

% At this point we would like to address the role of the driving $\tilde{a}^{\rm (in)}_{\rm P+}$ in the spectrum of eigenvalues. To do this we will compare the results of Sec.~\ref{sec:ActiveRing_Forced} with those obtained for the same active ring resonator where the driving has been switched off after triggering lasing in the $\ell_{\rm S}=133$ mode.

% First we look at the transmittance spectrum (Sec.~\ref{sec:ActiveRing_Unforced_Transmittance}) and then we carry a Bogoliubov analysis (Sec.~\ref{sec:ActiveRing_Unforced_Bogoliubov}).

% %
% %%%%%%%%%%%%%%%%%%%%%%%%%%%%%%%%%%%%%%%%%%%%%%%%%%%%
% \subsubsection{Transmittance}
% \label{sec:ActiveRing_Unforced_Transmittance}
% %%%%%%%%%%%%%%%%%%%%%%%%%%%%%%%%%%%%%%%%%%%%%%%%%%%%
% %

As before, we start by evaluating the  transmittance in the two directions for a weak signal beam around the resonance frequency $\omega^{(0)}_{\rm S}$ of the modes with $|\ell_{\rm S}|=131$. We use the same procedure that we followed in the presence of the seed: the results for $T_{\rm for,rev}$ shown in Fig.~\ref{fig:RingLaserDiode}d are practically identical to those obtained when the seed was switched on and shown in Fig.~\ref{fig:RingLaserDiode}b: This is not a surprise as the seed intensity is so small that it does not have any appreciable effect on the  nonlinear shift and the FWM coupling.

% %
% %%%%%%%%%%%%%%%%%%%%%%%%%%%%%%%%%%%%%%%%%%%%%%%%%%%%
% \subsubsection{Bogoliubov analysis}
% \label{sec:ActiveRing_Unforced_Bogoliubov}
% %%%%%%%%%%%%%%%%%%%%%%%%%%%%%%%%%%%%%%%%%%%%%%%%%%%%
% %

We then diagonalize the $4 \times 4$ Bogoliubov matrix in Eq.~\eqref{eq:BogoMatrixActive} to obtain the eigenvalues and eigenvectors. As expected, the values of the eigenvectors are only slightly modified by the absence of the seed,
\begin{align}
    &\|\mathbf{v}_{\rm I+}\|=-0.758,\;\;
    \|\mathbf{v}_{\rm S+}\|=0.758,\;\;
    \nonumber\\
    &\|\mathbf{v}_{\rm I-}\|=-1,\;\;
    \|\mathbf{v}_{\rm S-}\|=1.
\end{align}
The real parts of the eigenvalues $Re\{\omega_{\rm S}(\mathbf{v}_{\rm S,I\pm})\}$ are indicated as vertical lines in Fig.~\ref{fig:RingLaserDiode}d: once again, and exception made of course for $Re\{\omega_{\rm S}(\mathbf{v}_{\rm I-})\}$, the eigenvalues well match the position of the respective transmittance peaks.

The real and imaginary parts of the dispersion of $\omega_{\rm S}(\mathbf{v}_{\rm S,I\pm})$ are displayed in Fig.~\ref{fig:RingLaserDiode}e,f as a function of the angular momenta $|\ell_{\rm S}|$ in a neighborhood of $\ell_{\rm P}$. In the case of the reverse-propagating signal, very similar values of $\omega_{\rm S}(\mathbf{v}_{\rm S,I-})$ are found in the presence/absence of the seed. 
On the other hand, the eigenvalues $\omega_{\rm S}(\mathbf{v}_{\rm S,I+})$ corresponding to the forward case are significantly modified in a region of non-integer angular momenta around the pump angular momentum $\ell_{\rm P}$: their real parts coalesce at $Re\{\omega_{\rm S}(\mathbf{v}_{\rm S,I+})\}=\omega^{(0)}_{\rm P}-g_{\rm NL}|\tilde{a}^{(0)}_{\rm P+}|^2$ (i.e. the frequency of the Kerr-shifted resonance at $\ell_{\rm P}=133$), while simultaneously their imaginary parts split and take the different values $Im\{\omega_{\rm S}(\mathbf{v}_{\rm S+})\}=0$ and $Im\{\omega_{\rm S}(\mathbf{v}_{\rm I+})\}=-\gamma_{\rm T}$ at $\ell_{\rm S}=\ell_{\rm P}$. These correspond to the characteristic Goldstone and amplitude modes of a spontaneous U(1) symmetry breaking system~\cite{wouters2007goldstone}: here, the spontaneously-broken symmetry corresponds to the phase of the pump laser field amplitude $\tilde{a}^{(0)}_{\rm P+}$. In the presence of the seed, this symmetry was explicitly broken by the seed of amplitude $\tilde{a}^{\rm (in)}_{\rm P+}$, whose phase fixes the phase of the laser emission and makes a gap appear in the real part of the collective mode dispersion.

For the chosen parameters, the diffusive region where the imaginary parts $Im\{\omega_{\rm S}(\mathbf{v}_{\rm S,I+})\}$ split does not extend to the next mode, so the Goldstone physics can only be detected on the pump mode. A wider diffusive region could be obtained in devices displaying a weaker (in absolute value) curvature $\alpha$, so to access the Goldstone and amplitude modes also on the neighboring modes: as usual, the key signature of this phenomenon would be the narrowing (broadening) of the resonance peak corresponding to the Goldstone (amplitude) mode. Given the softness of the Goldstone mode, it is in any case beneficial to avoid operating the optical isolator in the diffusive region and, rather, choose a sufficiently distant signal mode. Such a choice has the further advantage of facilitating spectral rejection from the optical network downstream of the optical isolating element.

%
%%%%%%%%%%%%%%%%%%%%%%%%%%%%%%%%%%%%%%%%%%%%%%%%%%%%
\subsection{Connection with polariton condensates}
%%%%%%%%%%%%%%%%%%%%%%%%%%%%%%%%%%%%%%%%%%%%%%%%%%%%
%

While the coherently pumped ring resonator shared close analogies with a coherently pumped fluid of light, it is interesting to conclude this Section by highlighting the analogy with a non-equilibrium condensate or, if one prefers, a spatially extended laser device~\cite{Carusotto_2013,BCW2022}. In this case, the Bogoliubov modes probed by the signal correspond to the collective modes of the condensate which, in particular, feature a diffusive Goldstone branch at small wavevector and a gap opening in the presence of a coherent seed explicitly breaking the symmetry~\cite{wouters2007goldstone}. As condensation occurs in a mode with a well-defined chirality, the non-reciprocal optical isolator behavior can be again understood in terms of the dragging of collective excitations by the underlying moving fluid.

% is observed in the transmission of a weak signal probe when condensation is forced to occur in a moving state and collective excitations propagate differently in opposite directions: transferring this physical picture to our integrated photonic device, non-reciprocity and optical isolation arise in a lasing ring resonator as the collective excitations on top of the lasing mode are dragged by its chiral motion around the resonator. 

%%%%%%%%%%%%%%%%%%%%%%%%%%%%%%%%%%% TJR LASER-BASED OPTICAL ISOLATOR
%%%%%%%%%%%%%%%%%%%%%%%%%%%%%%%%%%

%
\begin{figure*}
    \centering
    \includegraphics[width=\textwidth]{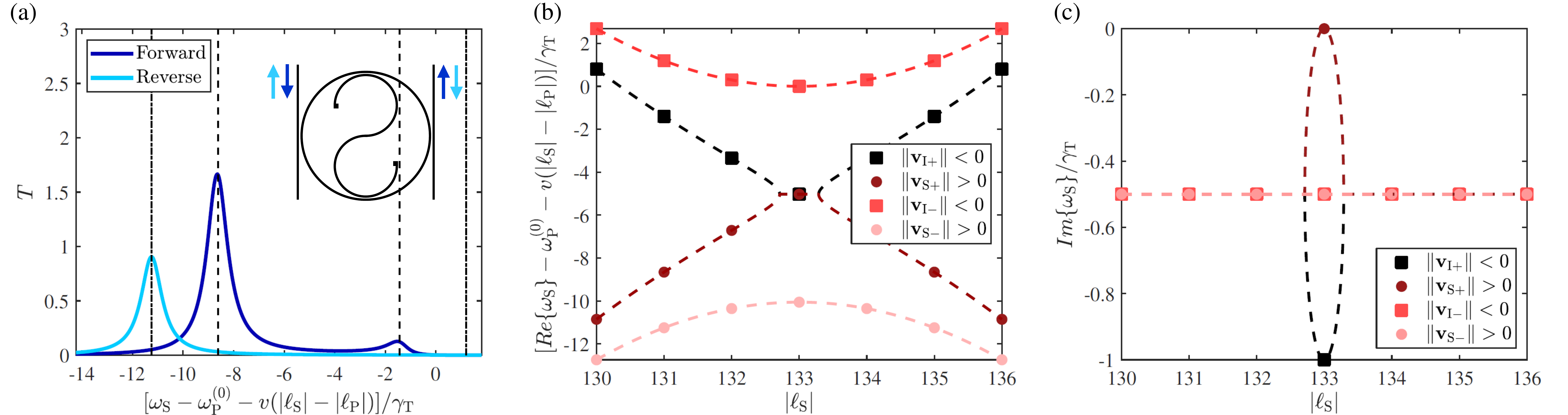}
    \caption{\textbf{(a)} Transmittance $T$ in forward (dark blue) and reverse (cyan) directions across an incoherently pumped active TJR in add-drop configuration lasing in the $\ell_{\rm P}=133$ mode, as a function of the signal frequency $\omega_{\rm S}$ for a pair of counterpropagating modes with angular momenta $|\ell_{\rm S}|=131$. The pump rate is fixed at $P_{0}=2\gamma_{\rm T}$. A sketch of the device is found in the upper-right part of the panel. The arrows describe the probing with the signal. The vertical dashed-dotted and dashed lines signal respectively the values of the real part of the Bogoliubov eigenvalues corresponding to the reverse and forward-propagating signals. \textbf{(b,c)} Real and imaginary parts of the Bogoliubov eigenvalues $\omega_{\rm S}$ as the angular momenta $\pm|\ell_{\rm S}|$ of the modes in which the counterpropagating signals are coupled is varied around $\pm\ell_{\rm P}=\pm 133$. Squares and circles are the values corresponding to the modes of integer angular momenta $|\ell_{\rm S}|$. The dashed lines are calculated for non-integer values of $|\ell_{\rm S}|$.}
    \label{fig:TaijiLaserDiode}
\end{figure*}

\section{Active TJR}
\label{sec:TaijiDiode_TJRlaser}

As an ultimate configuration, in this last Section we demonstrate how our scheme for non-reciprocity and optical isolation directly extends to the case of a TJR laser, with major practical advantages: as discussed in~\cite{MunozDeLasHeras_2021b}, active TJRs deterministically impose a well-defined chirality to the laser emission without the need for any external seed: through the intrinsic nonlinearity of saturable gain, the explicit breaking of spatial parity by the S-shaped element dynamically translates into an effective breaking of time-reversal symmetry in the lasing state which can only be stable in one direction and is then only weakly affected by all those back-scattering processes induced by fabrication imperfections.

Without loss of generality, we consider a TJR featuring an S waveguide that couples the CW into the CCW modes as sketched in the inset of Fig.~\ref{fig:TaijiLaserDiode}a. This results in stable lasing in the CCW direction, while lasing in the CW direction is destabilized~\cite{MunozDeLasHeras_2021b}. As before, we set a pump rate $P_{0}=2\gamma_{\rm T}$ which gives an intensity $|\tilde{a}^{(0)}_{\rm P +}|^2=n_{\rm S}$ circulating in the $\ell_{\rm P}=133$ mode only. We employ the usual absorption losses $k_{\rm A}=0.032$, the coupling parameters $k_{\rm w}=0.04$ and $k_{\rm S}=0.04$ (yielding a quality factor $Q\simeq 4\times 10^{4}$), and an effective nonlinear coefficient $g_{\rm NL}n_{\rm S}/\gamma_{\rm T}\simeq 5$. As in the previous Section, we start by analyzing the transmittance spectrum of small amplitude signals propagating in opposite directions through the TJR laser device and, then, we make use of a Bogoliubov analysis to shine physical light into the resulting non-reciprocity and optical isolation behavior.

%
%%%%%%%%%%%%%%%%%%%%%%%%%%%%%%%%%%%%%%%%%%%%%%%%%%%%
\subsection{Transmission spectra}
\label{sec:TJR_Transmittance}
%%%%%%%%%%%%%%%%%%%%%%%%%%%%%%%%%%%%%%%%%%%%%%%%%%%%
%

We first probe the system in the forward direction using a weak signal beam whose frequency $\omega_{\rm S}$ is varied around the resonance frequency $\omega^{(0)}_{\rm S}$ corresponding to the $\ell_{\rm S}=131$ mode. %The forward signal features an input intensity $|\tilde{a}^{\rm (in)}_{\rm S+}|^2$. 
As in the previous Section, the forward transmission spectrum $T_{\rm for}$ displayed as a dark blue line in Fig.~\ref{fig:TaijiLaserDiode}a features the characteristic doublet structure originating from FWM. Due to the absence of FWM, the analogous spectrum for the reverse direction shown as a cyan line
%We then probe the system in the reverse direction with a signal of the same incident intensity $|\tilde{a}^{\rm (in)}_{\rm S-}|^2=|\tilde{a}^{\rm (in)}_{\rm S+}|^2$ around the resonance frequency $\omega^{(0)}_{\rm S}$ of the $\ell_{\rm S}=-131$ mode. Similarly to what we obtained in Secs.~\ref{sec:TaijiDiode_PassiveRing} and~\ref{sec:TaijiDiode_ActiveRing}, $T_{\rm rev}$ is negligible over a broad frequency range and 
features a single peak at a frequency given by the Kerr nonlinearity shift of the signal mode resonance, different from the ones of the $T_{\rm for}$ doublet. In particular, at the frequency $[\omega_{\rm S}-\omega^{(0)}_{\rm P}-v(|\ell_{\rm S}|-|\ell_{\rm P}|)]=-8.51\gamma_{\rm T}$ where we find the maximum $T_{\rm for}$ we get a $17$ dB suppression of the $T_{\rm rev}$ signal.

As in all previous cases, the linearity of the coupled-mode equations~(\ref{eq:GeneralS}-\ref{eq:GeneralI}) ensures that forward- and reverse-propagating incident beams do not interact with each other, so the TJR laser can function as an effective optical isolator device. As already mentioned, the crucial advantage of the TJR configuration is that a given chirality for lasing is deterministically imposed by the laser operation with no need for seeding it. As a result, the active TJR can be considered as a stand-alone optical isolator device. A minor difference with respect to the previous cases, is that the S-shaped element introduces some finite reflection into the incident bus waveguide for light propagating in the reverse direction. While this feature does not appear to spoil the isolation behaviour, it might need to be taken into account in the design of the optical network that follows the optical isolator.

%
%%%%%%%%%%%%%%%%%%%%%%%%%%%%%%%%%%%%%%%%%%%%%%%%%%%%
\subsection{Bogoliubov analysis}
\label{sec:TJR_Bogoliubov}
%%%%%%%%%%%%%%%%%%%%%%%%%%%%%%%%%%%%%%%%%%%%%%%%%%%%
%

%Finally, we also studied the eigenvalues and eigenvectors of the Bogoliubov matrix coupling the forward and reverse signal and idler as a function of the angular momentum $|\ell_{\rm S}|$ of counterpropagating modes. 
In the TJR case the system of linearized differential equations features additional terms in the $4\times 4$ Bogoliubov matrix, which account for the coupling of the CW modes into the CCW ones enabled by the S waveguide of the TJR,
%\scriptsize
%
\begin{widetext}
\begin{align}
i & \frac{d}{dt}
\begin{bmatrix}
\tilde{a}_{\rm S+} &
\tilde{a}^{*}_{\rm I+} &
\tilde{a}_{\rm S-} &
\tilde{a}^{*}_{\rm I-}
\end{bmatrix}^{T}
\nonumber\\
& =
\begin{bmatrix}
%(1,1)
\begin{matrix}
\omega^{(0)}_{\rm S}-\omega_{\rm S}-i\gamma_{\rm T}\\
-2g_{\rm NL}|\tilde{a}^{(0)}_{\rm P+}|^2
\\
+i\frac{P_{0}}{1+\frac{1}{n_{\rm S}}|\tilde{a}^{(0)}_{\rm P+}|^2}
\\
-i\frac{P_{0}/n_{\rm S}}{\left[1+\frac{1}{n_{\rm S}}|\tilde{a}^{(0)}_{\rm P+}|^2\right]^2}|\tilde{a}^{(0)}_{\rm P+}|^2
\end{matrix}
&
%(1,2)
\begin{matrix}
-g_{\rm NL}\tilde{a}^{(0)^2}_{\rm P+}
\\
-i\frac{P_{0}/n_{\rm S}}{\left[1+\frac{1}{n_{\rm S}}|\tilde{a}_{\rm P+}|^2\right]^2}\tilde{a}^{(0)^2}_{\rm P+}
\end{matrix}
&
%(1,3)
\begin{matrix}
\beta^{\rm (S)}_{\pm}
\end{matrix}
&
%(1,4)
\begin{matrix}
0
\end{matrix}
\\
%(2,1)
\begin{matrix}
+g_{\rm NL}\tilde{a}^{(0)^{*2}}_{\rm P+}
\\
-i\frac{P_{0}/n_{\rm S}}{\left[1+\frac{1}{n_{\rm S}}|\tilde{a}^{(0)}_{\rm P+}|^2\right]^2}\tilde{a}^{(0)^{*2}}_{\rm P+}
\end{matrix}
&
%(2,2)
\begin{matrix}
-\omega^{(0)}_{\rm I}+2\omega_{\rm P}-\omega_{\rm S}-i\gamma_{\rm T}\\
+2g_{\rm NL}|\tilde{a}^{(0)}_{\rm P+}|^2
\\
+i\frac{P_{0}}{1+\frac{1}{n_{\rm S}}|\tilde{a}^{(0)}_{\rm P+}|^2}
\\
-i\frac{P_{0}/n_{\rm S}}{\left[1+\frac{1}{n_{\rm S}}|\tilde{a}^{(0)}_{\rm P+}|^2\right]^2}|\tilde{a}^{(0)}_{\rm P+}|^2
\end{matrix}
&
%(2,3)
\begin{matrix}
0
\end{matrix}
&
%(2,4)
\begin{matrix}
-\beta^{\rm (I) *}_{\pm}
\end{matrix}
\\
%(3,1)
\begin{matrix}
0
\end{matrix}
&
%(3,2)
\begin{matrix}
0
\end{matrix}
&
%(3,3)
\begin{matrix}
\omega^{(0)}_{\rm S}-\omega_{\rm S}-i\gamma_{\rm T}\\
-2g_{\rm NL}|\tilde{a}^{(0)}_{\rm P+}|^2
\\
+i\frac{P_{0}}{1+2\frac{1}{n_{\rm S}}|\tilde{a}^{(0)}_{\rm P+}|^2}
\\
-i\frac{P_{0}/n_{\rm S}}{\left[1+\frac{1}{n_{\rm S}}|\tilde{a}^{(0)}_{\rm P+}|^2\right]^2}|\tilde{a}^{(0)}_{\rm P+}|^2
\end{matrix}
&
%(3,4)
\begin{matrix}
0
\end{matrix}
\\
%(4,1)
\begin{matrix}
0
\end{matrix}
&
%(4,2)
\begin{matrix}
0
\end{matrix}
&
%(4,3)
\begin{matrix}
0
\end{matrix}
&
%(4,4)
\begin{matrix}
-\omega^{(0)}_{\rm I}+\omega_{\rm P}-\omega_{\rm S}-i\gamma_{\rm T}\\
+2g_{\rm NL}|\tilde{a}^{(0)}_{\rm P+}|^2
\\
+i\frac{P_{0}}{1+\frac{1}{n_{\rm S}}|\tilde{a}^{(0)}_{\rm P+}|^2}
\\
-i\frac{P_{0}/n_{\rm S}}{\left[1+\frac{1}{n_{\rm S}}|\tilde{a}^{(0)}_{\rm P+}|^2\right]^2}|\tilde{a}^{(0)}_{\rm P+}|^2
\end{matrix}
\end{bmatrix}
\nonumber\\
& \times
\begin{bmatrix}
\tilde{a}_{\rm S+} &
\tilde{a}^{*}_{\rm I+} &
\tilde{a}_{\rm S-} &
\tilde{a}^{*}_{\rm I-}
\end{bmatrix}^{T}
.
\label{eq:BogoMatrixTJR}
\end{align}
\end{widetext}
\normalsize
In spite of these additional terms, it follows from the theory of block matrices that the eigenvalues are not modified by the intermode $\beta^{\rm (S, I)}_{\pm}$ couplings and perfectly coincide with those of the active ring resonator in the absence of the external seed discussed around Eq.~\eqref{eq:BogoMatrixActive}. As expected, the position of the transmission peaks  perfectly match the real parts of these eigenvalues, indicated as vertical lines in Fig.~\ref{fig:TaijiLaserDiode}a.

In contrast, the eigenvectors $\mathbf{v}_{\rm S,I\pm}$ are significantly modified by the intermode couplings. In particular, the unidirectional coupling between the CW and CCW modes makes the Bogoliubov norms to be all different from $\pm 1$,
\begin{align}
    &\|\mathbf{v}_{\rm I+}\|=-0.756,\;\;
    \|\mathbf{v}_{\rm S+}\|=0.756,\;\;
    \nonumber\\
    &\|\mathbf{v}_{\rm I-}\|=-0.948,\;\;
    \|\mathbf{v}_{\rm S-}\|=0.948.
\end{align}
In spite of this, we still have no peak at $Re\{\omega_{\rm S}(\mathbf{v}_{\rm I-})\}$ in the reverse transmission due to the absence of FWM in the reverse direction. 

The real and imaginary parts of the corresponding eigenvalues $\omega_{\rm S}(\mathbf{v}_{\rm S,I\pm})$ are shown in Fig.~\ref{fig:TaijiLaserDiode}b,c as a function of $\ell_{\rm S}$: %To get a complete picture of the underlying physics, these eigenvalues are calculated also for non-integer values of $|\ell_{\rm S}|$. 
%As in the seed-less active ring resonator, 
the real parts $Re\{\omega_{\rm S}(\mathbf{v}_{\rm S,I-})\}$ corresponding to the reverse signal are distributed in two bands separated by a gap which achieves its minimum frequency width at $|\ell_{\rm S}|=\ell_{\rm P}=133$. $Re\{\omega_{\rm S}(\mathbf{v}_{\rm S-})\}$ determines the frequency at which the reverse transmittance $T_{\rm rev}$ takes its maximum value, which is again given by the resonance frequency of the $-\ell_{\rm S}$ mode shifted by the Kerr nonlinearity [see Eq.~\eqref{eq:ReOmega3}]. 
The corresponding imaginary parts take the same and negative value $Im\{\omega_{\rm S}(\mathbf{v}_{\rm S,I-})\}=-\gamma_{\rm T}/2$, confirming dynamical stability. 
%% On the other hand, no peak appears at the frequency given by $Re\{\omega_{\rm S}(\mathbf{v}_{\rm I-})\}$ due to the absence of FWM in the reverse direction. This is confirmed for the particular angular momentum $\ell_{\rm S}=-131$ in Fig.~\ref{fig:TaijiLaserDiode}a, where we have plotted both $Re\{\omega_{\rm S}(\mathbf{v}_{\rm S,I-})\}$ as vertical dashed-dotted lines. 

In the case of the forward-propagating signal, the two bands $Re\{\omega_{\rm S}(\mathbf{v}_{\rm S,I+})\}$ give again the frequencies of the transmittance doublet for each value of $\ell_{\rm S}$. As compared to the reverse signal case, the two bands are now closer between each other, implying that FWM produces an effective attraction between transmittance peaks. Furthermore, the real parts of $\omega_{\rm S}(\mathbf{v}_{\rm S,I+})$ coalesce in a non-integer neighborhood of $\ell_{\rm P}=133$, taking the value $Re\{\omega_{\rm S}(\mathbf{v}_{\rm S,I+})\}=\omega^{(0)}_{\rm P}-g_{\rm NL}|\tilde{a}^{(0)}_{\rm P+}|^2$ of the Kerr-shifted resonance at $\ell_{\rm P}=133$, and therefore closing the gap. In the same region, their corresponding imaginary parts $Im\{\omega_{\rm S}(\mathbf{v}_{\rm S,I+})\}$ depart from the value $-\gamma_{\rm T}/2$ and display the same Goldstone and amplitude modes discussed in detail in Sec.~\ref{sec:TaijiDiode_ActiveRing}.

%%%%%%%%%%%%%%%%%%%%%%%%%%%%%%%%%%% CONCLUSIONS
%%%%%%%%%%%%%%%%%%%%%%%%%%%%%%%%%%

\section{Conclusions}
\label{sec:Conclusions}

In this work, we have proposed and theoretically characterized a promising route to obtain efficient optical isolation in small-footprint integrated photonics devices. Our proposal takes inspiration from the dragging of the collective excitation modes in moving fluids of light~\cite{Carusotto_2013}: in particular, it does not require any explicit time-reversal-symmetry-breaking component such as magneto-optical elements, but instead exploits standard nonlinear optical processes occurring in generic dielectric materials in the presence of a strong unidirectional pump beam.

The proposed setup is based on a ring or Taiji resonator supporting a strong and coherent single-mode propagating field, which produces four-wave mixing processes in the forward direction only, and thus allows to overcome dynamic reciprocity restrictions~\cite{Shi_2015}. In this way, additional weak signals display a strongly non-reciprocal transmittance, such as a high transmission in the forward direction and a suppressed transmission in the reverse one (or viceversa). In combination with the efficient unidirectional lasing of Taiji resonators~\cite{MunozDeLasHeras_2021b}, our proposal then provides a complete platform for an efficient optical isolation component to be included in integrated photonic networks.

In addition to the experimental implementation of the proposed device, further theoretical research will address the interplay between our optical isolation mechanism and the nontrivial band topologies and topological edge states that can be engineered in arrays of ring and Taiji resonators and will explore its application to a new generation of spin-Hall topological laser devices~\cite{Harari_2018,Bandres_2018,Ozawa_2019,Ota_2020}.

%%%%%%%%%%%%%%%%%%%%%%%%%%%%%%%%%%% ACKNOWLEDGEMENTS
%%%%%%%%%%%%%%%%%%%%%%%%%%%%%%%%%%

\acknowledgements

We acknowledge financial support from the European Union FET-Open grant ``MIR-BOSE'' (n. 737017), from the H2020-FETFLAG-2018-2020 project ``PhoQuS'' (n.820392), from the Provincia Autonoma di Trento, and from the Q@TN initiative. A.M. acknowledges financial support from the Comunidad de Madrid via Proyecto Sinérgico CAM 2020 Y2020/TCS-6545 (NanoQuCo-CM). Stimulating discussions with Giuseppe La Rocca, Alberto Amo, Stefan Rotter, and Stefano Azzini are warmly acknowledged.

%%%%%%%%%%%%%%%%%%%%%%%%%%%%%%%
% Appendix
%%%%%%%%%%%%%%%%%%%%%%%%%%%%%%%

\appendix
\section{Partial differential equation for the field amplitude}
\label{sec:Appendix}

For the sake of completeness, it is useful to explicitly present the full partial differential equation for the field amplitude $a(\varphi,t)$, where $\varphi$ is the angular coordinate along the ring. Looking at this expression facilitates understanding the physical interpretation of our proposal in terms of a fluid of light circulating around the ring cavity.

Such an partial differential equation can be written as 
\begin{multline}
    i\frac{\partial a(\varphi,t)}{\partial t}=h_0\,a - g_{\rm NL} |a|^2 a + \\ + \frac{iP_0\,a}{1+|a|^2/n_{\rm S}} - i \gamma_{\rm T} a - F(\varphi,t) \label{eq:PDE}
\end{multline}
where $F(\varphi,t)$ encodes the external driving by the incident light and the linear evolution term $h_0 a$ is easiest written in angular-momentum space as 
\begin{multline}
    h_0 \bar{a}(\ell,t) = \omega_\ell^{(0)}\,\bar{a}(\ell,t)=\bigg[\omega^{(0)}_{\rm P}+v\left( |\ell|-|\ell_{\rm P}| \right)+
    \\
    +\frac{\alpha}{2}\left( |\ell|-|\ell_{\rm P}| \right)^2\bigg]\,\bar{a}(\ell,t)
\end{multline}
where $\bar{a}(\ell,t)=\frac{1}{2\pi}\int\!d\varphi\,a(\varphi,t)\,e^{-i\ell \varphi}$. In this formula, the roles of $v$ and $\alpha$ as respectively the group velocity and the inverse mass of photons circulating around the ring cavity are transparent. The intermode coupling terms proportional to the $\beta$ coefficients are instead easier introduced directly at the level of the coupled mode equations.

Based on this evolution equation, the equations of motion (\ref{eq:GeneralP}-\ref{eq:GeneralI}) for the $\ell_{\rm P},\pm\ell_{\rm S},\pm\ell_{\rm I}$ modes of interest can be obtained by expanding the field amplitude $a(\varphi,t)$ according to the Ansatz \eqref{eq:ansatz} and then projecting \eqref{eq:PDE} onto the relevant modes while only keeping first-order terms on the signal and idler field amplitudes.

\bibliography{Bibliography.bib}

\begin{thebibliography}{36}%
\makeatletter
\providecommand \@ifxundefined [1]{%
 \@ifx{#1\undefined}
}%
\providecommand \@ifnum [1]{%
 \ifnum #1\expandafter \@firstoftwo
 \else \expandafter \@secondoftwo
 \fi
}%
\providecommand \@ifx [1]{%
 \ifx #1\expandafter \@firstoftwo
 \else \expandafter \@secondoftwo
 \fi
}%
\providecommand \natexlab [1]{#1}%
\providecommand \enquote  [1]{``#1''}%
\providecommand \bibnamefont  [1]{#1}%
\providecommand \bibfnamefont [1]{#1}%
\providecommand \citenamefont [1]{#1}%
\providecommand \href@noop [0]{\@secondoftwo}%
\providecommand \href [0]{\begingroup \@sanitize@url \@href}%
\providecommand \@href[1]{\@@startlink{#1}\@@href}%
\providecommand \@@href[1]{\endgroup#1\@@endlink}%
\providecommand \@sanitize@url [0]{\catcode `\\12\catcode `\$12\catcode
  `\&12\catcode `\#12\catcode `\^12\catcode `\_12\catcode `\%12\relax}%
\providecommand \@@startlink[1]{}%
\providecommand \@@endlink[0]{}%
\providecommand \url  [0]{\begingroup\@sanitize@url \@url }%
\providecommand \@url [1]{\endgroup\@href {#1}{\urlprefix }}%
\providecommand \urlprefix  [0]{URL }%
\providecommand \Eprint [0]{\href }%
\providecommand \doibase [0]{https://doi.org/}%
\providecommand \selectlanguage [0]{\@gobble}%
\providecommand \bibinfo  [0]{\@secondoftwo}%
\providecommand \bibfield  [0]{\@secondoftwo}%
\providecommand \translation [1]{[#1]}%
\providecommand \BibitemOpen [0]{}%
\providecommand \bibitemStop [0]{}%
\providecommand \bibitemNoStop [0]{.\EOS\space}%
\providecommand \EOS [0]{\spacefactor3000\relax}%
\providecommand \BibitemShut  [1]{\csname bibitem#1\endcsname}%
\let\auto@bib@innerbib\@empty
%</preamble>
\bibitem [{\citenamefont {Jalas}\ \emph {et~al.}(2013)\citenamefont {Jalas},
  \citenamefont {Petrov}, \citenamefont {Eich}, \citenamefont {Freude},
  \citenamefont {Fan}, \citenamefont {Yu}, \citenamefont {Baets}, \citenamefont
  {Popovi{\'{c}}}, \citenamefont {Melloni}, \citenamefont {Joannopoulos},
  \citenamefont {Vanwolleghem}, \citenamefont {Doerr},\ and\ \citenamefont
  {Renner}}]{Jalas_2013}%
  \BibitemOpen
  \bibfield  {author} {\bibinfo {author} {\bibfnamefont {D.}~\bibnamefont
  {Jalas}}, \bibinfo {author} {\bibfnamefont {A.}~\bibnamefont {Petrov}},
  \bibinfo {author} {\bibfnamefont {M.}~\bibnamefont {Eich}}, \bibinfo {author}
  {\bibfnamefont {W.}~\bibnamefont {Freude}}, \bibinfo {author} {\bibfnamefont
  {S.}~\bibnamefont {Fan}}, \bibinfo {author} {\bibfnamefont {Z.}~\bibnamefont
  {Yu}}, \bibinfo {author} {\bibfnamefont {R.}~\bibnamefont {Baets}}, \bibinfo
  {author} {\bibfnamefont {M.}~\bibnamefont {Popovi{\'{c}}}}, \bibinfo {author}
  {\bibfnamefont {A.}~\bibnamefont {Melloni}}, \bibinfo {author} {\bibfnamefont
  {J.~D.}\ \bibnamefont {Joannopoulos}}, \bibinfo {author} {\bibfnamefont
  {M.}~\bibnamefont {Vanwolleghem}}, \bibinfo {author} {\bibfnamefont {C.~R.}\
  \bibnamefont {Doerr}},\ and\ \bibinfo {author} {\bibfnamefont
  {H.}~\bibnamefont {Renner}},\ }\bibfield  {title} {\bibinfo {title} {What is
  --- and what is not --- an optical isolator},\ }\href
  {https://doi.org/10.1038/nphoton.2013.185} {\bibfield  {journal} {\bibinfo
  {journal} {Nature Photonics}\ }\textbf {\bibinfo {volume} {7}},\ \bibinfo
  {pages} {579} (\bibinfo {year} {2013})}\BibitemShut {NoStop}%
\bibitem [{\citenamefont {Potton}(2004)}]{Potton_2004}%
  \BibitemOpen
  \bibfield  {author} {\bibinfo {author} {\bibfnamefont {R.~J.}\ \bibnamefont
  {Potton}},\ }\bibfield  {title} {\bibinfo {title} {Reciprocity in optics},\
  }\href {https://doi.org/10.1088/0034-4885/67/5/r03} {\bibfield  {journal}
  {\bibinfo  {journal} {Reports on Progress in Physics}\ }\textbf {\bibinfo
  {volume} {67}},\ \bibinfo {pages} {717} (\bibinfo {year} {2004})}\BibitemShut
  {NoStop}%
\bibitem [{\citenamefont {D\"{o}tsch}\ \emph {et~al.}(2005)\citenamefont
  {D\"{o}tsch}, \citenamefont {Bahlmann}, \citenamefont {Zhuromskyy},
  \citenamefont {Hammer}, \citenamefont {Wilkens}, \citenamefont {Gerhardt},
  \citenamefont {Hertel},\ and\ \citenamefont {Popkov}}]{Dotsch_2005}%
  \BibitemOpen
  \bibfield  {author} {\bibinfo {author} {\bibfnamefont {H.}~\bibnamefont
  {D\"{o}tsch}}, \bibinfo {author} {\bibfnamefont {N.}~\bibnamefont
  {Bahlmann}}, \bibinfo {author} {\bibfnamefont {O.}~\bibnamefont
  {Zhuromskyy}}, \bibinfo {author} {\bibfnamefont {M.}~\bibnamefont {Hammer}},
  \bibinfo {author} {\bibfnamefont {L.}~\bibnamefont {Wilkens}}, \bibinfo
  {author} {\bibfnamefont {R.}~\bibnamefont {Gerhardt}}, \bibinfo {author}
  {\bibfnamefont {P.}~\bibnamefont {Hertel}},\ and\ \bibinfo {author}
  {\bibfnamefont {A.~F.}\ \bibnamefont {Popkov}},\ }\bibfield  {title}
  {\bibinfo {title} {Applications of magneto-optical waveguides in integrated
  optics: review},\ }\href {https://doi.org/10.1364/JOSAB.22.000240} {\bibfield
   {journal} {\bibinfo  {journal} {J. Opt. Soc. Am. B}\ }\textbf {\bibinfo
  {volume} {22}},\ \bibinfo {pages} {240} (\bibinfo {year} {2005})}\BibitemShut
  {NoStop}%
\bibitem [{\citenamefont {Bi}\ \emph {et~al.}(2011)\citenamefont {Bi},
  \citenamefont {Hu}, \citenamefont {Jiang}, \citenamefont {Kim}, \citenamefont
  {Dionne}, \citenamefont {Kimerling},\ and\ \citenamefont {Ross}}]{Bi_2011}%
  \BibitemOpen
  \bibfield  {author} {\bibinfo {author} {\bibfnamefont {L.}~\bibnamefont
  {Bi}}, \bibinfo {author} {\bibfnamefont {J.}~\bibnamefont {Hu}}, \bibinfo
  {author} {\bibfnamefont {P.}~\bibnamefont {Jiang}}, \bibinfo {author}
  {\bibfnamefont {D.~H.}\ \bibnamefont {Kim}}, \bibinfo {author} {\bibfnamefont
  {G.~F.}\ \bibnamefont {Dionne}}, \bibinfo {author} {\bibfnamefont {L.~C.}\
  \bibnamefont {Kimerling}},\ and\ \bibinfo {author} {\bibfnamefont {C.~A.}\
  \bibnamefont {Ross}},\ }\bibfield  {title} {\bibinfo {title} {On-chip optical
  isolation in monolithically integrated non-reciprocal optical resonators},\
  }\href {https://doi.org/10.1038/nphoton.2011.270} {\bibfield  {journal}
  {\bibinfo  {journal} {Nature Photonics}\ }\textbf {\bibinfo {volume} {5}},\
  \bibinfo {pages} {758} (\bibinfo {year} {2011})}\BibitemShut {NoStop}%
\bibitem [{\citenamefont {Shoji}\ \emph {et~al.}(2012)\citenamefont {Shoji},
  \citenamefont {Ito}, \citenamefont {Shirato},\ and\ \citenamefont
  {Mizumoto}}]{Shoji_2012}%
  \BibitemOpen
  \bibfield  {author} {\bibinfo {author} {\bibfnamefont {Y.}~\bibnamefont
  {Shoji}}, \bibinfo {author} {\bibfnamefont {M.}~\bibnamefont {Ito}}, \bibinfo
  {author} {\bibfnamefont {Y.}~\bibnamefont {Shirato}},\ and\ \bibinfo {author}
  {\bibfnamefont {T.}~\bibnamefont {Mizumoto}},\ }\bibfield  {title} {\bibinfo
  {title} {Mzi optical isolator with si-wire waveguides by surface-activated
  direct bonding},\ }\href {https://doi.org/10.1364/OE.20.018440} {\bibfield
  {journal} {\bibinfo  {journal} {Opt. Express}\ }\textbf {\bibinfo {volume}
  {20}},\ \bibinfo {pages} {18440} (\bibinfo {year} {2012})}\BibitemShut
  {NoStop}%
\bibitem [{\citenamefont {Yan}\ \emph {et~al.}(2020)\citenamefont {Yan},
  \citenamefont {Yang}, \citenamefont {Liu}, \citenamefont {Zhang},
  \citenamefont {Xia}, \citenamefont {Kang}, \citenamefont {Yang},
  \citenamefont {Qin}, \citenamefont {Deng},\ and\ \citenamefont
  {Bi}}]{Yan_2020}%
  \BibitemOpen
  \bibfield  {author} {\bibinfo {author} {\bibfnamefont {W.}~\bibnamefont
  {Yan}}, \bibinfo {author} {\bibfnamefont {Y.}~\bibnamefont {Yang}}, \bibinfo
  {author} {\bibfnamefont {S.}~\bibnamefont {Liu}}, \bibinfo {author}
  {\bibfnamefont {Y.}~\bibnamefont {Zhang}}, \bibinfo {author} {\bibfnamefont
  {S.}~\bibnamefont {Xia}}, \bibinfo {author} {\bibfnamefont {T.}~\bibnamefont
  {Kang}}, \bibinfo {author} {\bibfnamefont {W.}~\bibnamefont {Yang}}, \bibinfo
  {author} {\bibfnamefont {J.}~\bibnamefont {Qin}}, \bibinfo {author}
  {\bibfnamefont {L.}~\bibnamefont {Deng}},\ and\ \bibinfo {author}
  {\bibfnamefont {L.}~\bibnamefont {Bi}},\ }\bibfield  {title} {\bibinfo
  {title} {Waveguide-integrated high-performance magneto-optical isolators and
  circulators on silicon nitride platforms},\ }\href
  {https://doi.org/10.1364/OPTICA.408458} {\bibfield  {journal} {\bibinfo
  {journal} {Optica}\ }\textbf {\bibinfo {volume} {7}},\ \bibinfo {pages}
  {1555} (\bibinfo {year} {2020})}\BibitemShut {NoStop}%
\bibitem [{\citenamefont {{Bhandare}}\ \emph {et~al.}(2005)\citenamefont
  {{Bhandare}}, \citenamefont {{Ibrahim}}, \citenamefont {{Sandel}},
  \citenamefont {{Hongbin Zhang}}, \citenamefont {{Wust}},\ and\ \citenamefont
  {{Noe}}}]{Bhandare05}%
  \BibitemOpen
  \bibfield  {author} {\bibinfo {author} {\bibfnamefont {S.}~\bibnamefont
  {{Bhandare}}}, \bibinfo {author} {\bibfnamefont {S.~K.}\ \bibnamefont
  {{Ibrahim}}}, \bibinfo {author} {\bibfnamefont {D.}~\bibnamefont {{Sandel}}},
  \bibinfo {author} {\bibnamefont {{Hongbin Zhang}}}, \bibinfo {author}
  {\bibfnamefont {F.}~\bibnamefont {{Wust}}},\ and\ \bibinfo {author}
  {\bibfnamefont {R.}~\bibnamefont {{Noe}}},\ }\bibfield  {title} {\bibinfo
  {title} {Novel nonmagnetic 30-db traveling-wave single-sideband optical
  isolator integrated in iii/v material},\ }\href
  {https://doi.org/10.1109/JSTQE.2005.845620} {\bibfield  {journal} {\bibinfo
  {journal} {IEEE Journal of Selected Topics in Quantum Electronics}\ }\textbf
  {\bibinfo {volume} {11}},\ \bibinfo {pages} {417} (\bibinfo {year}
  {2005})}\BibitemShut {NoStop}%
\bibitem [{\citenamefont {Yu}\ and\ \citenamefont
  {Fan}(2009)}]{yu2009complete}%
  \BibitemOpen
  \bibfield  {author} {\bibinfo {author} {\bibfnamefont {Z.}~\bibnamefont
  {Yu}}\ and\ \bibinfo {author} {\bibfnamefont {S.}~\bibnamefont {Fan}},\
  }\bibfield  {title} {\bibinfo {title} {Complete optical isolation created by
  indirect interband photonic transitions},\ }\href
  {https://doi.org/10.1038/nphoton.2008.273} {\bibfield  {journal} {\bibinfo
  {journal} {Nature Photonics}\ }\textbf {\bibinfo {volume} {3}},\ \bibinfo
  {pages} {91} (\bibinfo {year} {2009})}\BibitemShut {NoStop}%
\bibitem [{\citenamefont {Fang}\ \emph {et~al.}(2012)\citenamefont {Fang},
  \citenamefont {Yu},\ and\ \citenamefont {Fan}}]{Fang_2012}%
  \BibitemOpen
  \bibfield  {author} {\bibinfo {author} {\bibfnamefont {K.}~\bibnamefont
  {Fang}}, \bibinfo {author} {\bibfnamefont {Z.}~\bibnamefont {Yu}},\ and\
  \bibinfo {author} {\bibfnamefont {S.}~\bibnamefont {Fan}},\ }\bibfield
  {title} {\bibinfo {title} {Realizing effective magnetic field for photons by
  controlling the phase of dynamic modulation},\ }\href
  {https://doi.org/10.1038/nphoton.2012.236} {\bibfield  {journal} {\bibinfo
  {journal} {Nature Photonics}\ }\textbf {\bibinfo {volume} {6}},\ \bibinfo
  {pages} {782} (\bibinfo {year} {2012})}\BibitemShut {NoStop}%
\bibitem [{\citenamefont {Galland}\ \emph {et~al.}(2013)\citenamefont
  {Galland}, \citenamefont {Ding}, \citenamefont {Harris}, \citenamefont
  {Baehr-Jones},\ and\ \citenamefont {Hochberg}}]{Galland:13}%
  \BibitemOpen
  \bibfield  {author} {\bibinfo {author} {\bibfnamefont {C.}~\bibnamefont
  {Galland}}, \bibinfo {author} {\bibfnamefont {R.}~\bibnamefont {Ding}},
  \bibinfo {author} {\bibfnamefont {N.~C.}\ \bibnamefont {Harris}}, \bibinfo
  {author} {\bibfnamefont {T.}~\bibnamefont {Baehr-Jones}},\ and\ \bibinfo
  {author} {\bibfnamefont {M.}~\bibnamefont {Hochberg}},\ }\bibfield  {title}
  {\bibinfo {title} {Broadband on-chip optical non-reciprocity using phase
  modulators},\ }\href {https://doi.org/10.1364/OE.21.014500} {\bibfield
  {journal} {\bibinfo  {journal} {Opt. Express}\ }\textbf {\bibinfo {volume}
  {21}},\ \bibinfo {pages} {14500} (\bibinfo {year} {2013})}\BibitemShut
  {NoStop}%
\bibitem [{\citenamefont {Doerr}\ \emph {et~al.}(2014)\citenamefont {Doerr},
  \citenamefont {Chen},\ and\ \citenamefont {Vermeulen}}]{doerr2014silicon}%
  \BibitemOpen
  \bibfield  {author} {\bibinfo {author} {\bibfnamefont {C.~R.}\ \bibnamefont
  {Doerr}}, \bibinfo {author} {\bibfnamefont {L.}~\bibnamefont {Chen}},\ and\
  \bibinfo {author} {\bibfnamefont {D.}~\bibnamefont {Vermeulen}},\ }\bibfield
  {title} {\bibinfo {title} {Silicon photonics broadband modulation-based
  isolator},\ }\href {https://doi.org/10.1364/OE.22.004493} {\bibfield
  {journal} {\bibinfo  {journal} {Opt. Express}\ }\textbf {\bibinfo {volume}
  {22}},\ \bibinfo {pages} {4493} (\bibinfo {year} {2014})}\BibitemShut
  {NoStop}%
\bibitem [{\citenamefont {Fan}\ \emph {et~al.}(2012)\citenamefont {Fan},
  \citenamefont {Wang}, \citenamefont {Varghese}, \citenamefont {Shen},
  \citenamefont {Niu}, \citenamefont {Xuan}, \citenamefont {Weiner},\ and\
  \citenamefont {Qi}}]{Fan_2012}%
  \BibitemOpen
  \bibfield  {author} {\bibinfo {author} {\bibfnamefont {L.}~\bibnamefont
  {Fan}}, \bibinfo {author} {\bibfnamefont {J.}~\bibnamefont {Wang}}, \bibinfo
  {author} {\bibfnamefont {L.~T.}\ \bibnamefont {Varghese}}, \bibinfo {author}
  {\bibfnamefont {H.}~\bibnamefont {Shen}}, \bibinfo {author} {\bibfnamefont
  {B.}~\bibnamefont {Niu}}, \bibinfo {author} {\bibfnamefont {Y.}~\bibnamefont
  {Xuan}}, \bibinfo {author} {\bibfnamefont {A.~M.}\ \bibnamefont {Weiner}},\
  and\ \bibinfo {author} {\bibfnamefont {M.}~\bibnamefont {Qi}},\ }\bibfield
  {title} {\bibinfo {title} {An all-silicon passive optical diode},\ }\href
  {https://doi.org/10.1126/science.1214383} {\bibfield  {journal} {\bibinfo
  {journal} {Science}\ }\textbf {\bibinfo {volume} {335}},\ \bibinfo {pages}
  {447} (\bibinfo {year} {2012})}\BibitemShut {NoStop}%
\bibitem [{\citenamefont {Bender}\ \emph {et~al.}(2013)\citenamefont {Bender},
  \citenamefont {Factor}, \citenamefont {Bodyfelt}, \citenamefont {Ramezani},
  \citenamefont {Christodoulides}, \citenamefont {Ellis},\ and\ \citenamefont
  {Kottos}}]{Bender_2013}%
  \BibitemOpen
  \bibfield  {author} {\bibinfo {author} {\bibfnamefont {N.}~\bibnamefont
  {Bender}}, \bibinfo {author} {\bibfnamefont {S.}~\bibnamefont {Factor}},
  \bibinfo {author} {\bibfnamefont {J.~D.}\ \bibnamefont {Bodyfelt}}, \bibinfo
  {author} {\bibfnamefont {H.}~\bibnamefont {Ramezani}}, \bibinfo {author}
  {\bibfnamefont {D.~N.}\ \bibnamefont {Christodoulides}}, \bibinfo {author}
  {\bibfnamefont {F.~M.}\ \bibnamefont {Ellis}},\ and\ \bibinfo {author}
  {\bibfnamefont {T.}~\bibnamefont {Kottos}},\ }\bibfield  {title} {\bibinfo
  {title} {Observation of asymmetric transport in structures with active
  nonlinearities},\ }\href {https://doi.org/10.1103/PhysRevLett.110.234101}
  {\bibfield  {journal} {\bibinfo  {journal} {Phys. Rev. Lett.}\ }\textbf
  {\bibinfo {volume} {110}},\ \bibinfo {pages} {234101} (\bibinfo {year}
  {2013})}\BibitemShut {NoStop}%
\bibitem [{\citenamefont {Peng}\ \emph {et~al.}(2014)\citenamefont {Peng},
  \citenamefont {{\"O}zdemir}, \citenamefont {Lei}, \citenamefont {Monifi},
  \citenamefont {Gianfreda}, \citenamefont {Long}, \citenamefont {Fan},
  \citenamefont {Nori}, \citenamefont {Bender},\ and\ \citenamefont
  {Yang}}]{Peng_2014}%
  \BibitemOpen
  \bibfield  {author} {\bibinfo {author} {\bibfnamefont {B.}~\bibnamefont
  {Peng}}, \bibinfo {author} {\bibfnamefont {{\c{S}}.~K.}\ \bibnamefont
  {{\"O}zdemir}}, \bibinfo {author} {\bibfnamefont {F.}~\bibnamefont {Lei}},
  \bibinfo {author} {\bibfnamefont {F.}~\bibnamefont {Monifi}}, \bibinfo
  {author} {\bibfnamefont {M.}~\bibnamefont {Gianfreda}}, \bibinfo {author}
  {\bibfnamefont {G.~L.}\ \bibnamefont {Long}}, \bibinfo {author}
  {\bibfnamefont {S.}~\bibnamefont {Fan}}, \bibinfo {author} {\bibfnamefont
  {F.}~\bibnamefont {Nori}}, \bibinfo {author} {\bibfnamefont {C.~M.}\
  \bibnamefont {Bender}},\ and\ \bibinfo {author} {\bibfnamefont
  {L.}~\bibnamefont {Yang}},\ }\bibfield  {title} {\bibinfo {title}
  {Parity--time-symmetric whispering-gallery microcavities},\ }\href
  {https://doi.org/10.1038/nphys2927} {\bibfield  {journal} {\bibinfo
  {journal} {Nature Physics}\ }\textbf {\bibinfo {volume} {10}},\ \bibinfo
  {pages} {394} (\bibinfo {year} {2014})}\BibitemShut {NoStop}%
\bibitem [{\citenamefont {Chang}\ \emph {et~al.}(2014)\citenamefont {Chang},
  \citenamefont {Jiang}, \citenamefont {Hua}, \citenamefont {Yang},
  \citenamefont {Wen}, \citenamefont {Jiang}, \citenamefont {Li}, \citenamefont
  {Wang},\ and\ \citenamefont {Xiao}}]{Chang_2014}%
  \BibitemOpen
  \bibfield  {author} {\bibinfo {author} {\bibfnamefont {L.}~\bibnamefont
  {Chang}}, \bibinfo {author} {\bibfnamefont {X.}~\bibnamefont {Jiang}},
  \bibinfo {author} {\bibfnamefont {S.}~\bibnamefont {Hua}}, \bibinfo {author}
  {\bibfnamefont {C.}~\bibnamefont {Yang}}, \bibinfo {author} {\bibfnamefont
  {J.}~\bibnamefont {Wen}}, \bibinfo {author} {\bibfnamefont {L.}~\bibnamefont
  {Jiang}}, \bibinfo {author} {\bibfnamefont {G.}~\bibnamefont {Li}}, \bibinfo
  {author} {\bibfnamefont {G.}~\bibnamefont {Wang}},\ and\ \bibinfo {author}
  {\bibfnamefont {M.}~\bibnamefont {Xiao}},\ }\bibfield  {title} {\bibinfo
  {title} {Parity--time symmetry and variable optical isolation in
  active--passive-coupled microresonators},\ }\href
  {https://doi.org/10.1038/nphoton.2014.133} {\bibfield  {journal} {\bibinfo
  {journal} {Nature Photonics}\ }\textbf {\bibinfo {volume} {8}},\ \bibinfo
  {pages} {524} (\bibinfo {year} {2014})}\BibitemShut {NoStop}%
\bibitem [{\citenamefont {Mu\~noz de~las Heras}\ \emph
  {et~al.}(2021)\citenamefont {Mu\~noz de~las Heras}, \citenamefont {Franchi},
  \citenamefont {Biasi}, \citenamefont {Ghulinyan}, \citenamefont {Pavesi},\
  and\ \citenamefont {Carusotto}}]{MunozDeLasHeras_2021}%
  \BibitemOpen
  \bibfield  {author} {\bibinfo {author} {\bibfnamefont {A.}~\bibnamefont
  {Mu\~noz de~las Heras}}, \bibinfo {author} {\bibfnamefont {R.}~\bibnamefont
  {Franchi}}, \bibinfo {author} {\bibfnamefont {S.}~\bibnamefont {Biasi}},
  \bibinfo {author} {\bibfnamefont {M.}~\bibnamefont {Ghulinyan}}, \bibinfo
  {author} {\bibfnamefont {L.}~\bibnamefont {Pavesi}},\ and\ \bibinfo {author}
  {\bibfnamefont {I.}~\bibnamefont {Carusotto}},\ }\bibfield  {title} {\bibinfo
  {title} {Nonlinearity-induced reciprocity breaking in a single nonmagnetic
  taiji resonator},\ }\href {https://doi.org/10.1103/PhysRevApplied.15.054044}
  {\bibfield  {journal} {\bibinfo  {journal} {Phys. Rev. Applied}\ }\textbf
  {\bibinfo {volume} {15}},\ \bibinfo {pages} {054044} (\bibinfo {year}
  {2021})}\BibitemShut {NoStop}%
\bibitem [{\citenamefont {Shi}\ \emph {et~al.}(2015)\citenamefont {Shi},
  \citenamefont {Yu},\ and\ \citenamefont {Fan}}]{Shi_2015}%
  \BibitemOpen
  \bibfield  {author} {\bibinfo {author} {\bibfnamefont {Y.}~\bibnamefont
  {Shi}}, \bibinfo {author} {\bibfnamefont {Z.}~\bibnamefont {Yu}},\ and\
  \bibinfo {author} {\bibfnamefont {S.}~\bibnamefont {Fan}},\ }\bibfield
  {title} {\bibinfo {title} {Limitations of nonlinear optical isolators due to
  dynamic reciprocity},\ }\href {https://doi.org/10.1038/nphoton.2015.79}
  {\bibfield  {journal} {\bibinfo  {journal} {Nature Photonics}\ }\textbf
  {\bibinfo {volume} {9}},\ \bibinfo {pages} {388} (\bibinfo {year}
  {2015})}\BibitemShut {NoStop}%
\bibitem [{\citenamefont {Bino}\ \emph {et~al.}(2018)\citenamefont {Bino},
  \citenamefont {Silver}, \citenamefont {Woodley}, \citenamefont {Stebbings},
  \citenamefont {Zhao},\ and\ \citenamefont {Del'Haye}}]{DelBino_2018}%
  \BibitemOpen
  \bibfield  {author} {\bibinfo {author} {\bibfnamefont {L.~D.}\ \bibnamefont
  {Bino}}, \bibinfo {author} {\bibfnamefont {J.~M.}\ \bibnamefont {Silver}},
  \bibinfo {author} {\bibfnamefont {M.~T.~M.}\ \bibnamefont {Woodley}},
  \bibinfo {author} {\bibfnamefont {S.~L.}\ \bibnamefont {Stebbings}}, \bibinfo
  {author} {\bibfnamefont {X.}~\bibnamefont {Zhao}},\ and\ \bibinfo {author}
  {\bibfnamefont {P.}~\bibnamefont {Del'Haye}},\ }\bibfield  {title} {\bibinfo
  {title} {Microresonator isolators and circulators based on the intrinsic
  nonreciprocity of the kerr effect},\ }\href
  {https://doi.org/10.1364/OPTICA.5.000279} {\bibfield  {journal} {\bibinfo
  {journal} {Optica}\ }\textbf {\bibinfo {volume} {5}},\ \bibinfo {pages} {279}
  (\bibinfo {year} {2018})}\BibitemShut {NoStop}%
\bibitem [{\citenamefont {Mu\~noz de~las Heras}\ and\ \citenamefont
  {Carusotto}(2021)}]{MunozDeLasHeras_2021b}%
  \BibitemOpen
  \bibfield  {author} {\bibinfo {author} {\bibfnamefont {A.}~\bibnamefont
  {Mu\~noz de~las Heras}}\ and\ \bibinfo {author} {\bibfnamefont
  {I.}~\bibnamefont {Carusotto}},\ }\bibfield  {title} {\bibinfo {title}
  {Unidirectional lasing in nonlinear taiji microring resonators},\ }\href
  {https://doi.org/10.1103/PhysRevA.104.043501} {\bibfield  {journal} {\bibinfo
   {journal} {Phys. Rev. A}\ }\textbf {\bibinfo {volume} {104}},\ \bibinfo
  {pages} {043501} (\bibinfo {year} {2021})}\BibitemShut {NoStop}%
\bibitem [{\citenamefont {Hohimer}\ \emph {et~al.}(1993)\citenamefont
  {Hohimer}, \citenamefont {Vawter},\ and\ \citenamefont
  {Craft}}]{Hohimer_1993}%
  \BibitemOpen
  \bibfield  {author} {\bibinfo {author} {\bibfnamefont {J.~P.}\ \bibnamefont
  {Hohimer}}, \bibinfo {author} {\bibfnamefont {G.~A.}\ \bibnamefont
  {Vawter}},\ and\ \bibinfo {author} {\bibfnamefont {D.~C.}\ \bibnamefont
  {Craft}},\ }\bibfield  {title} {\bibinfo {title} {Unidirectional operation in
  a semiconductor ring diode laser},\ }\href {https://doi.org/10.1063/1.108728}
  {\bibfield  {journal} {\bibinfo  {journal} {Applied Physics Letters}\
  }\textbf {\bibinfo {volume} {62}},\ \bibinfo {pages} {1185} (\bibinfo {year}
  {1993})}\BibitemShut {NoStop}%
\bibitem [{\citenamefont {Hohimer}\ and\ \citenamefont
  {Vawter}(1993)}]{Hohimer_1993b}%
  \BibitemOpen
  \bibfield  {author} {\bibinfo {author} {\bibfnamefont {J.~P.}\ \bibnamefont
  {Hohimer}}\ and\ \bibinfo {author} {\bibfnamefont {G.~A.}\ \bibnamefont
  {Vawter}},\ }\bibfield  {title} {\bibinfo {title} {Unidirectional
  semiconductor ring lasers with racetrack cavities},\ }\href
  {https://doi.org/10.1063/1.110474} {\bibfield  {journal} {\bibinfo  {journal}
  {Applied Physics Letters}\ }\textbf {\bibinfo {volume} {63}},\ \bibinfo
  {pages} {2457} (\bibinfo {year} {1993})}\BibitemShut {NoStop}%
\bibitem [{\citenamefont {Carusotto}\ and\ \citenamefont
  {Ciuti}(2013)}]{Carusotto_2013}%
  \BibitemOpen
  \bibfield  {author} {\bibinfo {author} {\bibfnamefont {I.}~\bibnamefont
  {Carusotto}}\ and\ \bibinfo {author} {\bibfnamefont {C.}~\bibnamefont
  {Ciuti}},\ }\bibfield  {title} {\bibinfo {title} {\textit{Quantum fluids of
  light}},\ }\href {https://doi.org/10.1103/RevModPhys.85.299} {\bibfield
  {journal} {\bibinfo  {journal} {Rev. Mod. Phys.}\ }\textbf {\bibinfo {volume}
  {85}},\ \bibinfo {pages} {299} (\bibinfo {year} {2013})}\BibitemShut
  {NoStop}%
\bibitem [{\citenamefont {Butcher}\ and\ \citenamefont
  {Cotter}(1990)}]{butcher_cotter_1990}%
  \BibitemOpen
  \bibfield  {author} {\bibinfo {author} {\bibfnamefont {P.~N.}\ \bibnamefont
  {Butcher}}\ and\ \bibinfo {author} {\bibfnamefont {D.}~\bibnamefont
  {Cotter}},\ }\href {https://doi.org/10.1017/CBO9781139167994} {\emph
  {\bibinfo {title} {The Elements of Nonlinear Optics}}},\ Cambridge Studies in
  Modern Optics\ (\bibinfo  {publisher} {Cambridge University Press},\ \bibinfo
  {year} {1990})\BibitemShut {NoStop}%
\bibitem [{\citenamefont {Walls}\ and\ \citenamefont
  {Milburn}(1994)}]{Walls1994_InputOutput}%
  \BibitemOpen
  \bibfield  {author} {\bibinfo {author} {\bibfnamefont {D.~F.}\ \bibnamefont
  {Walls}}\ and\ \bibinfo {author} {\bibfnamefont {G.~J.}\ \bibnamefont
  {Milburn}},\ }\bibinfo {title} {Input-output formulation of optical
  cavities},\ in\ \href {https://doi.org/10.1007/978-3-642-79504-6_7} {\emph
  {\bibinfo {booktitle} {Quantum Optics}}}\ (\bibinfo  {publisher} {Springer
  Berlin Heidelberg},\ \bibinfo {address} {Berlin, Heidelberg},\ \bibinfo
  {year} {1994})\ pp.\ \bibinfo {pages} {121--135}\BibitemShut {NoStop}%
\bibitem [{\citenamefont {Trenti}\ \emph {et~al.}(2018)\citenamefont {Trenti},
  \citenamefont {Borghi}, \citenamefont {Biasi}, \citenamefont {Ghulinyan},
  \citenamefont {Ramiro-Manzano}, \citenamefont {Pucker},\ and\ \citenamefont
  {Pavesi}}]{Trenti_2018}%
  \BibitemOpen
  \bibfield  {author} {\bibinfo {author} {\bibfnamefont {A.}~\bibnamefont
  {Trenti}}, \bibinfo {author} {\bibfnamefont {M.}~\bibnamefont {Borghi}},
  \bibinfo {author} {\bibfnamefont {S.}~\bibnamefont {Biasi}}, \bibinfo
  {author} {\bibfnamefont {M.}~\bibnamefont {Ghulinyan}}, \bibinfo {author}
  {\bibfnamefont {F.}~\bibnamefont {Ramiro-Manzano}}, \bibinfo {author}
  {\bibfnamefont {G.}~\bibnamefont {Pucker}},\ and\ \bibinfo {author}
  {\bibfnamefont {L.}~\bibnamefont {Pavesi}},\ }\bibfield  {title} {\bibinfo
  {title} {Thermo-optic coefficient and nonlinear refractive index of silicon
  oxynitride waveguides},\ }\href {https://doi.org/10.1063/1.5018016}
  {\bibfield  {journal} {\bibinfo  {journal} {AIP Advances}\ }\textbf {\bibinfo
  {volume} {8}},\ \bibinfo {pages} {025311} (\bibinfo {year}
  {2018})}\BibitemShut {NoStop}%
\bibitem [{\citenamefont {Hambenne}\ and\ \citenamefont
  {Sargent}(1975)}]{Hambenne_1975}%
  \BibitemOpen
  \bibfield  {author} {\bibinfo {author} {\bibfnamefont {J.}~\bibnamefont
  {Hambenne}}\ and\ \bibinfo {author} {\bibfnamefont {M.}~\bibnamefont
  {Sargent}},\ }\bibfield  {title} {\bibinfo {title} {Physical interpretation
  of bistable unidirectional ring-laser operation},\ }\href
  {https://doi.org/10.1109/JQE.1975.1068559} {\bibfield  {journal} {\bibinfo
  {journal} {IEEE Journal of Quantum Electronics}\ }\textbf {\bibinfo {volume}
  {11}},\ \bibinfo {pages} {90} (\bibinfo {year} {1975})}\BibitemShut {NoStop}%
\bibitem [{Note1()}]{Note1}%
  \BibitemOpen
  \bibinfo {note} {In these equations, we have neglected for simplicity the
  (typically small) frequency dependence of the nonlinearity and of the
  radiative couplings. Under this approximation, we have taken constant values
  for $g_{\protect \rm NL}=n_{\protect \rm NL}\omega ^{(0)}_{\protect \rm
  P}/n_{\protect \rm L}$ and for all $\gamma $'s regardless of the considered
  mode.}\BibitemShut {Stop}%
\bibitem [{\citenamefont {Mu\~noz de~las Heras}\ \emph
  {et~al.}(2020)\citenamefont {Mu\~noz de~las Heras}, \citenamefont
  {Macaluso},\ and\ \citenamefont {Carusotto}}]{MunozDeLasHeras_2020}%
  \BibitemOpen
  \bibfield  {author} {\bibinfo {author} {\bibfnamefont {A.}~\bibnamefont
  {Mu\~noz de~las Heras}}, \bibinfo {author} {\bibfnamefont {E.}~\bibnamefont
  {Macaluso}},\ and\ \bibinfo {author} {\bibfnamefont {I.}~\bibnamefont
  {Carusotto}},\ }\bibfield  {title} {\bibinfo {title} {Anyonic molecules in
  atomic fractional quantum hall liquids: A quantitative probe of fractional
  charge and anyonic statistics},\ }\href
  {https://doi.org/10.1103/PhysRevX.10.041058} {\bibfield  {journal} {\bibinfo
  {journal} {Phys. Rev. X}\ }\textbf {\bibinfo {volume} {10}},\ \bibinfo
  {pages} {041058} (\bibinfo {year} {2020})}\BibitemShut {NoStop}%
\bibitem [{\citenamefont {Gibbs}\ \emph {et~al.}(1976)\citenamefont {Gibbs},
  \citenamefont {McCall},\ and\ \citenamefont {Venkatesan}}]{Gibbs_1976}%
  \BibitemOpen
  \bibfield  {author} {\bibinfo {author} {\bibfnamefont {H.~M.}\ \bibnamefont
  {Gibbs}}, \bibinfo {author} {\bibfnamefont {S.~L.}\ \bibnamefont {McCall}},\
  and\ \bibinfo {author} {\bibfnamefont {T.~N.~C.}\ \bibnamefont
  {Venkatesan}},\ }\bibfield  {title} {\bibinfo {title} {Differential gain and
  bistability using a sodium-filled fabry-perot interferometer},\ }\href
  {https://doi.org/10.1103/PhysRevLett.36.1135} {\bibfield  {journal} {\bibinfo
   {journal} {Phys. Rev. Lett.}\ }\textbf {\bibinfo {volume} {36}},\ \bibinfo
  {pages} {1135} (\bibinfo {year} {1976})}\BibitemShut {NoStop}%
\bibitem [{\citenamefont {Claude}\ \emph {et~al.}(2021)\citenamefont {Claude},
  \citenamefont {Jacquet}, \citenamefont {Usciati}, \citenamefont {Carusotto},
  \citenamefont {Giacobino}, \citenamefont {Bramati},\ and\ \citenamefont
  {Glorieux}}]{claude2021highresolution}%
  \BibitemOpen
  \bibfield  {author} {\bibinfo {author} {\bibfnamefont {F.}~\bibnamefont
  {Claude}}, \bibinfo {author} {\bibfnamefont {M.~J.}\ \bibnamefont {Jacquet}},
  \bibinfo {author} {\bibfnamefont {R.}~\bibnamefont {Usciati}}, \bibinfo
  {author} {\bibfnamefont {I.}~\bibnamefont {Carusotto}}, \bibinfo {author}
  {\bibfnamefont {E.}~\bibnamefont {Giacobino}}, \bibinfo {author}
  {\bibfnamefont {A.}~\bibnamefont {Bramati}},\ and\ \bibinfo {author}
  {\bibfnamefont {Q.}~\bibnamefont {Glorieux}},\ }\href@noop {} {\bibinfo
  {title} {High-resolution coherent probe spectroscopy of a polariton quantum
  fluid}} (\bibinfo {year} {2021}),\ \Eprint {https://arxiv.org/abs/2112.09903}
  {arXiv:2112.09903 [cond-mat.quant-gas]} \BibitemShut {NoStop}%
\bibitem [{\citenamefont {Wouters}\ and\ \citenamefont
  {Carusotto}(2007)}]{wouters2007goldstone}%
  \BibitemOpen
  \bibfield  {author} {\bibinfo {author} {\bibfnamefont {M.}~\bibnamefont
  {Wouters}}\ and\ \bibinfo {author} {\bibfnamefont {I.}~\bibnamefont
  {Carusotto}},\ }\bibfield  {title} {\bibinfo {title} {Goldstone mode of
  optical parametric oscillators in planar semiconductor microcavities in the
  strong-coupling regime},\ }\href@noop {} {\bibfield  {journal} {\bibinfo
  {journal} {Physical Review A}\ }\textbf {\bibinfo {volume} {76}},\ \bibinfo
  {pages} {043807} (\bibinfo {year} {2007})}\BibitemShut {NoStop}%
\bibitem [{\citenamefont {Bloch}\ \emph {et~al.}(2021)\citenamefont {Bloch},
  \citenamefont {Carusotto},\ and\ \citenamefont {Wouters}}]{BCW2022}%
  \BibitemOpen
  \bibfield  {author} {\bibinfo {author} {\bibfnamefont {J.}~\bibnamefont
  {Bloch}}, \bibinfo {author} {\bibfnamefont {I.}~\bibnamefont {Carusotto}},\
  and\ \bibinfo {author} {\bibfnamefont {M.}~\bibnamefont {Wouters}},\ }\href
  {https://doi.org/10.48550/ARXIV.2106.11137} {\bibinfo {title} {Spontaneous
  coherence in spatially extended photonic systems: Non-equilibrium
  bose-einstein condensation}} (\bibinfo {year} {2021})\BibitemShut {NoStop}%
\bibitem [{\citenamefont {Harari}\ \emph {et~al.}(2018)\citenamefont {Harari},
  \citenamefont {Bandres}, \citenamefont {Lumer}, \citenamefont {Rechtsman},
  \citenamefont {Chong}, \citenamefont {Khajavikhan}, \citenamefont
  {Christodoulides},\ and\ \citenamefont {Segev}}]{Harari_2018}%
  \BibitemOpen
  \bibfield  {author} {\bibinfo {author} {\bibfnamefont {G.}~\bibnamefont
  {Harari}}, \bibinfo {author} {\bibfnamefont {M.~A.}\ \bibnamefont {Bandres}},
  \bibinfo {author} {\bibfnamefont {Y.}~\bibnamefont {Lumer}}, \bibinfo
  {author} {\bibfnamefont {M.~C.}\ \bibnamefont {Rechtsman}}, \bibinfo {author}
  {\bibfnamefont {Y.~D.}\ \bibnamefont {Chong}}, \bibinfo {author}
  {\bibfnamefont {M.}~\bibnamefont {Khajavikhan}}, \bibinfo {author}
  {\bibfnamefont {D.~N.}\ \bibnamefont {Christodoulides}},\ and\ \bibinfo
  {author} {\bibfnamefont {M.}~\bibnamefont {Segev}},\ }\bibfield  {title}
  {\bibinfo {title} {Topological insulator laser: Theory},\ }\href
  {https://science.sciencemag.org/content/359/6381/eaar4003} {\bibfield
  {journal} {\bibinfo  {journal} {Science}\ }\textbf {\bibinfo {volume} {359}}
  (\bibinfo {year} {2018})}\BibitemShut {NoStop}%
\bibitem [{\citenamefont {Bandres}\ \emph {et~al.}(2018)\citenamefont
  {Bandres}, \citenamefont {Wittek}, \citenamefont {Harari}, \citenamefont
  {Parto}, \citenamefont {Ren}, \citenamefont {Segev}, \citenamefont
  {Christodoulides},\ and\ \citenamefont {Khajavikhan}}]{Bandres_2018}%
  \BibitemOpen
  \bibfield  {author} {\bibinfo {author} {\bibfnamefont {M.~A.}\ \bibnamefont
  {Bandres}}, \bibinfo {author} {\bibfnamefont {S.}~\bibnamefont {Wittek}},
  \bibinfo {author} {\bibfnamefont {G.}~\bibnamefont {Harari}}, \bibinfo
  {author} {\bibfnamefont {M.}~\bibnamefont {Parto}}, \bibinfo {author}
  {\bibfnamefont {J.}~\bibnamefont {Ren}}, \bibinfo {author} {\bibfnamefont
  {M.}~\bibnamefont {Segev}}, \bibinfo {author} {\bibfnamefont {D.~N.}\
  \bibnamefont {Christodoulides}},\ and\ \bibinfo {author} {\bibfnamefont
  {M.}~\bibnamefont {Khajavikhan}},\ }\bibfield  {title} {\bibinfo {title}
  {Topological insulator laser: Experiments},\ }\href
  {https://science.sciencemag.org/content/359/6381/eaar4005} {\bibfield
  {journal} {\bibinfo  {journal} {Science}\ }\textbf {\bibinfo {volume} {359}}
  (\bibinfo {year} {2018})}\BibitemShut {NoStop}%
\bibitem [{\citenamefont {Ozawa}\ \emph {et~al.}(2019)\citenamefont {Ozawa},
  \citenamefont {Price}, \citenamefont {Amo}, \citenamefont {Goldman},
  \citenamefont {Hafezi}, \citenamefont {Lu}, \citenamefont {Rechtsman},
  \citenamefont {Schuster}, \citenamefont {Simon}, \citenamefont {Zilberberg},\
  and\ \citenamefont {Carusotto}}]{Ozawa_2019}%
  \BibitemOpen
  \bibfield  {author} {\bibinfo {author} {\bibfnamefont {T.}~\bibnamefont
  {Ozawa}}, \bibinfo {author} {\bibfnamefont {H.~M.}\ \bibnamefont {Price}},
  \bibinfo {author} {\bibfnamefont {A.}~\bibnamefont {Amo}}, \bibinfo {author}
  {\bibfnamefont {N.}~\bibnamefont {Goldman}}, \bibinfo {author} {\bibfnamefont
  {M.}~\bibnamefont {Hafezi}}, \bibinfo {author} {\bibfnamefont
  {L.}~\bibnamefont {Lu}}, \bibinfo {author} {\bibfnamefont {M.~C.}\
  \bibnamefont {Rechtsman}}, \bibinfo {author} {\bibfnamefont {D.}~\bibnamefont
  {Schuster}}, \bibinfo {author} {\bibfnamefont {J.}~\bibnamefont {Simon}},
  \bibinfo {author} {\bibfnamefont {O.}~\bibnamefont {Zilberberg}},\ and\
  \bibinfo {author} {\bibfnamefont {I.}~\bibnamefont {Carusotto}},\ }\bibfield
  {title} {\bibinfo {title} {\textit{Topological photonics}},\ }\href
  {https://doi.org/10.1103/RevModPhys.91.015006} {\bibfield  {journal}
  {\bibinfo  {journal} {Rev. Mod. Phys.}\ }\textbf {\bibinfo {volume} {91}},\
  \bibinfo {pages} {015006} (\bibinfo {year} {2019})}\BibitemShut {NoStop}%
\bibitem [{\citenamefont {Ota}\ \emph {et~al.}(2020)\citenamefont {Ota},
  \citenamefont {Takata}, \citenamefont {Ozawa}, \citenamefont {Amo},
  \citenamefont {Jia}, \citenamefont {Kante}, \citenamefont {Notomi},
  \citenamefont {Arakawa},\ and\ \citenamefont {Iwamoto}}]{Ota_2020}%
  \BibitemOpen
  \bibfield  {author} {\bibinfo {author} {\bibfnamefont {Y.}~\bibnamefont
  {Ota}}, \bibinfo {author} {\bibfnamefont {K.}~\bibnamefont {Takata}},
  \bibinfo {author} {\bibfnamefont {T.}~\bibnamefont {Ozawa}}, \bibinfo
  {author} {\bibfnamefont {A.}~\bibnamefont {Amo}}, \bibinfo {author}
  {\bibfnamefont {Z.}~\bibnamefont {Jia}}, \bibinfo {author} {\bibfnamefont
  {B.}~\bibnamefont {Kante}}, \bibinfo {author} {\bibfnamefont
  {M.}~\bibnamefont {Notomi}}, \bibinfo {author} {\bibfnamefont
  {Y.}~\bibnamefont {Arakawa}},\ and\ \bibinfo {author} {\bibfnamefont
  {S.}~\bibnamefont {Iwamoto}},\ }\bibfield  {title} {\bibinfo {title} {Active
  topological photonics},\ }\href
  {https://doi.org/doi:10.1515/nanoph-2019-0376} {\bibfield  {journal}
  {\bibinfo  {journal} {Nanophotonics}\ }\textbf {\bibinfo {volume} {9}},\
  \bibinfo {pages} {547} (\bibinfo {year} {2020})}\BibitemShut {NoStop}%
\end{thebibliography}%

\end{document}